\documentclass[12pt,a4paper]{article}

\usepackage{amssymb,amsmath,amsthm,amsfonts,eurosym,geometry,caption,color,setspace,comment,footmisc,caption,apacite,natbib,pdflscape,array,hyperref,bbm,dcolumn,multicol,graphicx,subcaption,adjustbox,lipsum,apalike,booktabs,cleveref,cases,float,appendix,blindtext,comment,lscape,threeparttable}

\graphicspath{{"G:/My Drive/TSE/PhD/Third paper/Draft/arXiv"}}

\excludecomment{hidden}

\newcolumntype{L}[1]{>{\raggedright\arraybackslash}m{#1}}
\newcolumntype{C}[1]{>{\centering\arraybackslash}m{#1}}
\newcolumntype{R}[1]{>{\raggedleft\arraybackslash}m{#1}}

\floatstyle{plaintop}
\restylefloat{table}

\renewcommand{\footnoterule}{\vfill\kern -3pt \hrule width 0.4\columnwidth \kern 2.6pt}

\begin{document}
	

	\begin{titlepage}

		\title{\vspace{-3cm}Power to the teens? \\ A model of parents' and teens' collective labor supply } 
		\date{\today}
		\author{Jos\'e Alfonso Mu\~noz-Alvarado\thanks{Email: \href{mailto:jalfonso.munoza@gmail.com}{jalfonso.munoza@gmail.com}. I like to thank Fran\c{c}ois Poinas, Thierry Magnac, Olivier de Groote, Bram de Rock, and Matteo Bobba for their support and guidance on this paper. I would like to also thank the participants in different workshops at TSE and Université Libre de Bruxelles.} \\ Toulouse School of Economics}

		\maketitle
		
		\begin{abstract}
			\noindent Teens make life-changing decisions while constrained by the needs and resources of the households they grow up in. Household behavior models frequently delegate decision-making to the teen or their parents, ignoring joint decision-making in the household. I show that teens and parents allocate time and income jointly by using data from the Costa Rican \textit{Encuesta Nacional de Hogares} from 2011 to 2019 and a conditional cash transfer program. First, I present gender differences in household responses to the transfer using a marginal treatment effect framework. Second, I explain how the gender gap from the results is due to the bargaining process between parents and teens. I propose a collective household model and show that sons bargain cooperatively with their parents while daughters do not. This result implies that sons have a higher opportunity cost of attending school than daughters. Public policy targeting teens must account for this gender disparity to be effective.
			\bigskip
		\end{abstract}
		\smallskip
		\noindent \textbf{Keywords:}  Structural modeling, collective household, gender, youth, marginal treatment effect. \\
		\noindent \textbf{JEL Classification:}  D11, D12, D13, C31, J13, J22
		\setcounter{page}{0}
		\thispagestyle{empty}
		
	\end{titlepage}
	\newpage

	\doublespacing
	
\section{Introduction}

Teens make decisions that have lifelong implications for their human capital accumulation, potential income, and welfare \citep{cunha2006interpreting}. These decisions are made along with their household's needs, such as housework or earning a wage, and may result in them dropping out of school. It raises concern since school dropouts earn less and have fewer work possibilities for the rest of their lives \citep{de2016out}. To avoid this, governments use subsidies such as conditional cash transfers (CCT), which give cash incentives to households as long as the teens continue to attend school. In addition to their benefits on educational outcomes in primary and secondary schools \citep{garcia2017educational}, CCTs have significant effects on households.

CCTs' have a direct impact on households by relaxing their budget constraints. \cite{todd2006assessing} provide evidence that the Mexican CCT \textit{Oportunidades} influences parental fertility decisions via this effect. Their unitary model assumes that households are single decision-making agents. However, this model does not account for the effect of CCTs on intrahousehold bargaining. To quantify this effect, a collective model of household decision-making in which multiple decision-makers allocate income, time, and consumption \citep{chiappori1992collective} is needed. \cite{de2022household} show how \textit{Oportunidades} changes household consumption patterns, including those of young children, using such a model. This result is partly due to the mother’s greater bargaining power and increased sharing with her children. Although teens have been included in collective models of consumption, the literature has ignored the possibility that they allocate time cooperatively with their parents, thereby overlooking the teens' preferences and opportunity costs.

In this paper, I present evidence that teens allocate time by bargaining cooperatively with their parents. Using the Costa Rican \textit{Encuesta Nacional de Hogares} (ENAHO) from 2011 to 2019, I provide two sets of results. First, I present reduced-form evidence that a CCT program designed to keep teens in high school changes household time outcomes. Using a marginal treatment effect approach that accounts for the CCT program's endogeneity, I show that teens' schooling and parents' labor outcomes differ by gender for households that are indifferent between applying or not applying to the transfer program. If the teen is a daughter, her parents enjoy more leisure by decreasing their labor outcomes because of the program. Sons increase their school attendance, but only their mothers enjoy more leisure time, not their fathers. These gender differences in household responses to the CCT may be related to teens' bargaining power in the household. The second set of results investigates whether the teen bargains collectively with her parents about time allocation. I present a collective household model where teens and parents bargain to distribute income and time. Among other empirical results, I find that daughter does not cooperatively bargain with her parents. The son does, however, and is considered a decision-maker. I provide estimates of the bargaining function through which households with a teen son allocate resources. To the best of my knowledge, this is one of the first papers to consider the teen decision-maker in a time-use framework.

For my analysis, I evaluate the Costa Rican conditional cash transfer \textit{Avancemos} on household decision-making. This CCT was established in 2006 and provides cash transfers to households as long as the teen remains in high school. The data I use comes from the Costa Rican ENAHO, a yearly household survey that collects individual data on a representative sample of households. It includes demographic, domestic, and labor market variables, and whether the household receives \textit{Avancemos}. For the reduced-form results, I estimate a marginal treatment effect as presented by \cite{heckman2005structural}. As an instrument, I use the percentage of households that have \textit{Avancemos} in the neighborhood. This variable is a proxy for treatment take-up spillover effects, accounting for the stigma, or willingness, associated with receiving social transfers and does not affect household outcomes. The instrument can be considered exogenous because the peer effect is on treatment take-up rather than on outcomes.

To rationalize gender differences in household responses to \textit{Avancemos}, I present a collective household model in which households make decisions about teen high school participation, the labor supply of the parents, and domestic public good production. This model is built on \cite{blundell2007collective} and \cite{cherchye2012married}. I use the model predicted restrictions to find whether the teen is a decision-maker. The primary goal of the test is to compare how changes in wages and nonlabor income impact the outcomes of the parents and teens. In a unitary model with a single agent, transfers and earned wages have the same income effect on the household because they relax the budget constraint. The collective model with two or more decision-makers has an additional effect generated by transfers and wages because of the bargaining process. Aside from the income effect, a decision-maker who brings extra income to the household gains more gains bargaining power. My results show that although daughters are more effective at housework than sons, they do not have a bargaining position in the household. Because sons do negotiate with their parents, I can recover their bargaining function. The sons’ share of income rises with additional transfers and their wages. However, they only receive a portion of their parents' wages, which are the primary source of income for the household. Last, as a result of the bargaining, the sons' income share decreases if they attend school but do not work.

These findings are useful in policy design. First, the opportunity cost of attending school for sons includes the bargaining effect, whereas it does not for daughters. Furthermore, daughters complement their education with more domestic work, which decreases their opportunity cost. Policies targeting teens must account for this gender difference to be effective. In addition, these effects may result in sons specializing in labor and daughters in domestic work. This specialization can have longer term effects on lifetime earnings and human capital accumulation as shown by women having lower labor participation even with higher education than men.  As a result, public policies aimed at closing gender gaps should take gender roles into account from an early age in households.

The paper is structured as follows: In the next section, I discuss the literature. Section three gives \textit{Avancemos}' background. The fourth section presents the data and empirical evidence on the impact of \textit{Avancemos} on household time allocation decisions. I continue with the theoretical model in section five and empirical specifications of the unitary and collective models in section six. Results are shown in the seventh section. Section eight concludes.

\section{Related literature}

My paper adds first and foremost to the collective household model literature. \citet{chiappori1992collective} presents a study of household consumption and labor supply as a collective decision between the parents of a household. \citet{chiappori2017static} provide a comprehensive review of the literature. The literature concentrated on parents as decision-makers until the study by \citet{dauphin2011children}. Their results provide compelling evidence that teens who are sixteen years old or older cooperatively bargain with their parents about consumption allocation. This result, together with the method presented by \cite{dunbar2013children}, extended the literature by incorporating children as decision-makers. These analyses, however, only include consumption decisions and not labor supply. This is explained by the difficulty of observing children's wages and an exogenous variation in the nonlabor income of teens to identify their share of income as decision-makers. I contribute to the literature by developing a collective model and conducting empirical estimations with a teen as a decision-maker in a time-use framework using a rich dataset from Costa Rica and the CCT \textit{Avancemos}.

My paper also builds on the literature on teen education decision-making. In a household setting, parents make decisions on behalf of the teen. \cite{del2014household}, for example, study parents’ labor supply and the cognitive development process of children. They find that cash transfers to households with children have small impacts on child quality production because a significant fraction is spent on other household consumption and the leisure of the parents. Few articles investigate teens' schooling decisions in a collective model. \cite{reggio2011influence} and \citet{keshavarz2017parents}, for example, adopt a collective model where mothers decide their children's schooling or work decisions. However, \cite{lundberg2009decision} argue that teens become capable of productive and independent work while continuing to depend on their parents. \cite{ashraf2020negotiating} provide evidence that teens can negotiate their education. My paper is the first to use a collective bargaining approach in which teens make time-allocation decisions alongside their parents.

The paper adds to the literature on the effects of transfers in households. For example, \citet{attanasio2014efficient} show how the mother, the recipient of the transfer in \textit{Oportunidades}, gains more control over household resources, affecting household consumption and labor market behavior. \cite{dunbar2013children} propose an estimation that allows the determination of the share of household resources for children. A large body of literature follows this approach, including \citet{brown2018sharing}, \cite{tommasi2019control}, \citet{calvi2020older} and \citet{calvi2021more}. They extend the analysis by developing a measure of intrahousehold poverty.
Furthermore, \citet{sokullu2022individual} analyze the effect of \textit{Oportunidades} and suggest a novel identification strategy for recovering the household share of consumption, including children. In a similar topic, but without employing a collective model, \cite{dubois2012child} shows how CCTs make households reallocate care time for younger children between the mother and eldest daughter. I contribute to the literature by focusing on a transfer's effect on intrahousehold bargaining over the allocation of teens' time.

\section{Institutional background}
By 2012, Costa Rica had a secondary school dropout rate of 10\%, higher than its primary school rate of 2.5\% \citep{mata2015evaluacion}. An explanation for this difference is higher opportunity costs of sending children to school rather than working in households with tighter budgets. In 2006, the Costa Rican government implemented \textit{Avancemos}, a conditional cash transfer program to retain teens from low-income households in secondary education \citep{munoz2016avancemos}. Every public high school offers the possibility of applying to it. Once a household applies, \textit{Avancemos} is assigned according to the Social Information Sheet (FIS) of the Mixed Institute for Social Aid (IMAS) \citep{mata2015evaluacion}. The FIS collects household information and assigns a poverty score based on the wealth and demographics of the household. The household can apply for \textit{Avancemos} if the teen is between the ages of 12 and 25 years old and is registered in high school\footnote{Costa Rican high school system consists of 5 years, where the usual age for the student is from 13 to 17 years old. However, there are night high schools for older students who work during the day.} \citep{imasweb}. Households that receive \textit{Avancemos} must send the teen to at least 80\% of monthly classes. The amount of \textit{Avancemos} initially varied across the five years of high school, increasing each year. Since 2015, the amount has been fixed at 42 US dollars\footnote{The exchange rate used was the average of the selling exchange rate indicated by the Central Bank of Costa Rica for the first 8 months of 2015: 540.77 colones per US dollar} per month for the first three years of high school and 65 US dollars per month for the last two years\footnote{Technical high schools in Costa Rica have an extra year and the amount of \textit{Avancemos} is the same as the year before.} \citep{hernandez2016como}. In comparison, the monthly cost of a basic food basket in Costa Rica in July 2015 was 84 US dollars \citep{enaho2015}.

Some papers have investigated the effects of \textit{Avancemos}. \cite{meza2011effects} shows that it increases beneficiaries' education by half a year, two-thirds of a year for men. \cite{mata2015evaluacion} find an increase between 10 and 16\% in high school retention due to the program. \textit{Avancemos} also decreased child labor by approximately two percent between 2005 and 2006, with a greater effect on rural areas for male students \citep{lang2015efecto}. There have been no studies on the impact of \textit{Avancemos} on parents' outcomes. For other children in the household,  \cite{munoz2016avancemos} finds no effect on the likelihood of beneficiary siblings' attending school. Common across these results are gender differences, with male students benefiting more than female students. One explanation is the difference in the opportunity cost of schooling between daughters and sons. According to \cite{jimenez2015analysis}, women specialize in domestic work while men focus on market work in Costa Rica. This specialization may go hand in hand with schooling and domestic work being more complementary, lowering the opportunity cost for daughters to attend school compared with sons. My collective model includes this potential mechanism by modeling schooling, work in the labor market, and domestic time.

\section{Data}
I use the ENAHO repeated cross-section data from 2011 to 2019 \citep{enaho}. It consists of a representative sample of Costa Rican households, such as nuclear families, single parents without children, extended families, and so on. It collects information on the demographic, labor market, and domestic variables of each household member. It also includes information about the household, such as transfers and who receives \textit{Avancemos}. The survey is ideal for my analysis for two reasons. First, the survey collects labor variables regardless of whether the person works in the formal or informal sector. It accomplishes this by assuring respondents that their information will not be shared with tax authorities, as well as by structuring the labor-related questions so that they do not directly ask about the individual's sector of employment. This characteristic allows me to observe teens' labor variables and wages, as they are more likely to be in the informal sector due to their low-skill level. Second, the ENAHO collects the number of hours people dedicate to housework (cleaning, chores, caring for younger children and the elderly).

The ENAHO sample consists of 100,989 households, with an average of 11,000 households each year. I select nuclear families (mother, father, and children) whose parents are aged 30 to 64, whose oldest child is between ages 15 and 20 years old, and with any other children 14 years old or younger. The main reason for this selection is to have only one child who is legally allowed to work\footnote{15 years old is the legal age to start working.} and no other adult in the household who can provide extra income.  I also choose households in which the father works and eliminate observations with extreme values on wages, labor hours, and total household income\footnote{All monetary variables are corrected for inflation with the base year being 2019. Additionally, I normalized the units to refer to one thousand colones, corresponding to 1.58 US dollars.}.

Table \ref{tab:household} shows descriptive statistics of my sample. Most families have more than one child, are not poor\footnote{Define in the ENAHO under a poverty line that assesses the household's ability to meet its needs through consumption.}, and live in a rural area outside the Central Valley or in an urban area in the Central Valley.
\begin{table}[H]
	\centering
	\caption{Households' descriptive statistics}\label{tab:household} \singlespacing
	\begin{tabular}{rccccc}  \hline \hline
		\multicolumn{1}{c}{\textbf{Variable}} & \textbf{Obs} & \textbf{Mean} & \textbf{Std. Dev.} & \textbf{Min} & \textbf{Max} \\
		\midrule
		Total household income & 2,447 & 163.67 & 77.62 & 30.31 & 480.61 \\
		Poor  & 2,447 & 0.20  &      
		& 0     & 1 \\
		Not poor & 2,447 & 0.80  &       & 0     & 1 \\
		\textit{\textbf{Number of children}} &       &       &       &       & 
		\\
		One   & 2,447 & 0.30  &      
		& 0     & 1 \\
		Two   & 2,447 & 0.42  &      
		& 0     & 1 \\
		Three or more & 2,447 & 0.28 
		&       & 0     & 1 \\
		\textit{\textbf{Geographic urban area}} &       &       &       &       & 
		\\
		Outside Central Valley, rural area & 2,447 & 0.35  &      
		& 0     & 1 \\
		Central Valley, rural area & 2,447 & 0.21 
		&       & 0     & 1 \\
		Outside Central Valley, urban area & 2,447 & 0.17  &      
		& 0     & 1 \\
		Central Valley, urban area & 2,447 & 0.27 &       & 0     & 1 \\
		\hline \hline
	\end{tabular}%
\end{table}%

Table \ref{tab:parents/desc} shows descriptive statistics of parents. Most of the parents have a school degree. There is a significant difference between fathers and mothers in the labor market and domestic hours. Mothers work an average of 40 hours per week in the home, while fathers work an average of 3.5 hours per week. However, fathers work an average of 55 hours per week in the labor market, while 34\% of mothers have a job and work an average of 38 hours per week.

\begin{table}[H]
	\centering
	\caption{Parents' descriptive statistics}\label{tab:parents/desc} \singlespacing
	\begin{adjustbox}{width=1.1\textwidth}
		\begin{threeparttable}
			\begin{tabular}{rccccccccccc}  \hline \hline
				& \multicolumn{5}{c}{\textbf{Father}}   &       & \multicolumn{5}{c}{\textbf{Mother}} \\
				\cmidrule{2-6}\cmidrule{8-12}    \multicolumn{1}{c}{\textbf{Variable}} & \textbf{Obs} & \textbf{Mean} & \textbf{Std. Dev.} & \textbf{Min} & \textbf{Max} &       & \textbf{Obs} & \textbf{Mean} & \textbf{Std. Dev.} & \textbf{Min} & \textbf{Max} \\
				\cmidrule{1-6}\cmidrule{8-12}          &       &       &       &       &       &       &       &       &       &       &  \\
				Age   & 2,447 & 44.57 & 7.13  & 30.00 & 64.00 &       & 2,447 & 40.61 & 6.10  & 30.00 & 62.00 \\
				Primary School diploma & 2,447 & 0.66  &       & 0.00  & 1.00  &       & 2,447 & 0.66  &       & 0.00  & 1.00 \\
				High School diploma or more & 2,447 & 0.17  &       & 0.00  & 1.00  &       & 2,447 & 0.17  &       & 0.00  & 1.00 \\
				Years of Schooling & 2,447 & 6.96  & 3.15  & 0.00  & 17.00 &       & 2,447 & 7.14  & 3.14  & 0.00  & 17.00 \\
				&       &       &       &       &       &       &       &       &       &       &  \\
				Employed & 2,447 & 1.00  &       & 1.00  & 1.00  &       & 2,447 & 0.31  & 0.46  & 0.00  & 1.00 \\
				Hourly wage* & 2,447 & 1.78  & 0.83  & 0.17  & 4.62  &       & 770   & 1.66  & 0.90  & 0.10  & 4.70 \\
				Market hours* & 2,447 & 49.86 & 11.93 & 8.00  & 81.00 &       & 770   & 38.37 & 15.52 & 6.00  & 80.00 \\
				Domestic participation & 2,447 & 0.47  &       & 0.00  & 1.00  &       & 2,447 & 0.98  &       & 0.00  & 1.00 \\
				Domestic hours* & 1,157 & 7.27  & 6.06  & 1.00  & 35.00 &       & 2,408 & 40.30 & 19.32 & 1.00  & 89.00 \\
				\hline \hline
			\end{tabular}%
			\begin{tablenotes}\footnotesize
				\item[] *: Conditional on participation.
			\end{tablenotes}
		\end{threeparttable}
	\end{adjustbox}
\end{table}%

Table \ref{tab:teens/desc} shows a clear gender difference in school attendance and domestic work supply for teens. Daughters are more likely than sons to help out around the house (86 percent vs. 62 percent), and if they do, they put in almost twice as many hours (11 hours per week vs 6 hours per week). On the other hand, sons are more likely to work and support the family financially and are less likely to attend high school.

\begin{table}[H]
	\centering
	\caption{Teens' descriptive statistics}\label{tab:teens/desc} \singlespacing
	\begin{adjustbox}{width=1.1\textwidth}
		\begin{threeparttable}
			\begin{tabular}{rcccccrccccc}\hline \hline
				& \multicolumn{5}{c}{\textbf{Sons}}     &       & \multicolumn{5}{c}{\textbf{Daughters}} \\
				\cmidrule{2-6}\cmidrule{8-12}    \multicolumn{1}{c}{\textbf{Variable}} & \textbf{Obs} & \textbf{Mean} & \textbf{Std. Dev.} & \textbf{Min} & \textbf{Max} &       & \textbf{Obs} & \textbf{Mean} & \textbf{Std. Dev.} & \textbf{Min} & \textbf{Max} \\
				\cmidrule{1-6}\cmidrule{8-12}          &       &       &       &       &       &       &       &       &       &       &  \\
				Age   & 1,357 & 16.68 & 1.46  & 15.00 & 20.00 &       & 1,090 & 16.39 & 1.32  & 15.00 & 20.00 \\
				Age$\geq$18 & 1,357 & 0.27  &       & 0.00  & 1.00  &       & 1,090 & 0.19  &       & 0.00  & 1.00 \\
				Decision &       &       &       &       &       &       &       &       &       &       &  \\
				Nothing & 1,357 & 0.13  &       & 0.00  & 1.00  &       & 1,090 & 0.11  &       & 0.00  & 1.00 \\
				School & 1,357 & 0.75  &       & 0.00  & 1.00  &       & 1,090 & 0.86  &       & 0.00  & 1.00 \\
				Paid work & 1,357 & 0.13  &       & 0.00  & 1.00  &       & 1,090 & 0.02  &       & 0.00  & 1.00 \\
				&       &       &       &       &       &       &       &       &       &       &  \\
				Avancemos & 1,357 & 0.28  &       & 0.00  & 1.00  &       & 1,090 & 0.35  &       & 0.00  & 1.00 \\
				Hourly wage* & 170   & 1.26  & 0.59  & 0.20  & 3.92  &       & 26    & 1.17  & 0.39  & 0.35  & 1.99 \\
				Market hours* & 170   & 44.85 & 14.58 & 8.00  & 78.00 &       & 26    & 40.62 & 16.48 & 13.00 & 80.00 \\
				Domestic participation & 1,357 & 0.625 &       & 0     & 1     &       & 1,090 & 0.858 &       & 0     & 1 \\
				Domestic hours* & 847   & 6.795 & 5.613 & 1     & 29    &       & 935   & 11.450 & 9.138 & 1     & 42 \\
				\hline \hline
			\end{tabular}%
			\begin{tablenotes}\footnotesize
				\item[] *: Conditional on participation.
			\end{tablenotes}
		\end{threeparttable}
	\end{adjustbox}
\end{table}%

\subsection{Marginal Treatment Effect of \textit{Avancemos}}
To analyze the effect of \textit{Avancemos}, I define the treated households as those who receive the transfer. In my sample, 70\% of the 762 households that were treated had parents who only have a high school diploma, and 80\% of those households have at least one other younger child. This suggests an endogenous treatment selection, where households with higher costs of sending their teen to school are those that receive \textit{Avancemos}. I cannot control for treatment selection because I do not see which households apply for it or not in my data, only the households that benefit from it. To deal with this selection into treatment take-up, I estimate the marginal treatment effect developed by \cite{heckman2005structural}\footnote{\cite{blundell2009alternative} and \cite{cornelissen2016late} provide a survey on the marginal treatment effect, its extensions, and applications.}. From a policy point of view, this estimator enables us to know the benefits of \textit{Avancemos} of households in the margin to apply or not by estimating its marginal costs and benefits. Knowing this makes it easier to adjust the subsidy to the policy's objectives and limitations.

Consider a household $i$ that can receive the binary treatment of \textit{Avancemos}
($D_{i}=1$) or not ($D_{i}=0$) and has a vector of covariates $X_{i}$. Its observed outcome $Y_{i}$ is linked to the potential outcomes through the following equation:
\begin{equation*}
	Y_{i} = (1-D_{i})Y_{0i} + D_{i}Y_{1i}
\end{equation*}
The potential outcomes are specified as follows:
\begin{equation*}
	Y_{ji} = \mu_{j}(X_{i}) + U_{ji} \; \; j = 0,1,\\
\end{equation*}
where $\mu_{j}(\cdot)$ are unspecified functions and  $U_{ji}$ are random variables with mean zero conditional on covariates. The latent variable discrete choice model for selection into treatment is:
\begin{align*}
	D_{i}^{*} &= \mu_{D}(X_{i}, Z_{i}) - V_{i} \\
	D_{i} &= 1, \; \; \textrm{if} \; \; D_{i}^{*}\geq0,  \; \; \; \; D_{i} = 0, \; \; \textrm{otherwise}
\end{align*}
where $Z_{i}$ is a vector of instruments and $V_{i}$ is an i.i.d error denoting unobserved heterogeneity in the propensity for treatment. Because $V_{i}$ enters the selection equation with a negative sign, it can be thought of as an unobserved distaste for being treated. Assuming $F$ as the cumulative distribution function for $V$, the propensity score function is defined as $P(X_{i},Z_{i}) = F_{V}(\mu_{D}(X_{i}, Z_{i}))$, and $U_{Di} = F_{V}(V_{i})$ represents the quantiles of the distribution of the unobserved distaste for treatment. Rewriting the latent variable model for selection into treatment as:
\begin{equation*}
	D_{i} = 1, \; \; \textrm{if} \; \; P(X_{i},Z_{i})\geq U_{Di},
\end{equation*}
The marginal treatment effect on the household's outcomes of receiving \textit{Avancemos} for households with observables $X_{i}=x$ and unobserved distaste of treatment $U_{Di}=p$ is:
\begin{equation*}\label{eq:mte}
	\textrm{MTE}(x,p) = E[Y_{1i} - Y_{0i}| X = x, U_{D} = p] = \mu_{1}(x) - \mu_{0}(x) + E[U_{1} - U_{0}| X = x, U_{D} = p].
\end{equation*}
The values $p$ correspond to the values of the propensity score to receive \textit{Avancemos}. I estimate the MTE where there is common support across the propensity score. The MTE can be interpreted intuitively as the average effect of treatment for households on the margin of indifference between applying or not applying to \textit{Avancemos}. The indifference margin is set by the values of the propensity score. The propensity score values determine the indifference margin. The instrument's use allows for an exogenous variation, in which the change in treatment status captures the treatment effect for similar values of the propensity score. The similarities in the assumptions required to estimate the MTE and LATE estimators are explained by \cite{cornelissen2016late}. 

\subsubsection{Instrument}
Instruments that are related to distance are frequently used in literature. \cite{carneiro2011estimating}, for example, use distance to the nearest high school to estimate the effect of educational returns on future wages. I am unable to create a distance-related variable with my data, but I can calculate the proportion of neighborhood households that receive \textit{Avancemos}  to get a proxy of the peer effect\footnote{\cite{dipasquale1999incentives} uses a similar instrument in a home ownership framework.}. I use the entire ENAHO sample of 100,989 households to create this variable. I define 112 geographic blocks annually, each with an average of 100 households, and estimate the number of households in each household's block, excluding itself, that benefit from \textit{Avancemos}. In my sample, households live in neighborhoods where on average 10\% of households benefit from \textit{Avancemos}.

For this instrument to be valid, it must fulfill two conditions: (i) The instrument $Z_{i}$ is a random variable such that the propensity score $P(X_{i},Z_{i})$ is a nontrivial function of $Z_{i}$; and (ii) $(U_{0i} ,U_{1i},V_{i}) $ are independent of $Z_{i}$, conditional on $X_{i}$. For condition (i), the instrument serves as a proxy for the household's positive or negative peer effects. If a middle-class household is the only one in the neighborhood to have the transfer, it might feel socially stigmatized as a result of getting it. A lower-income household can easily learn about and apply for \textit{Avancemos}, though, if a lot of the homes in the neighborhood are receiving it. Table \ref{tab:instruments} shows that the instrument is relevant and has an impact on the uptake of the treatment.

\begin{table}[H]\centering
	\def\sym#1{\ifmmode^{#1}\else\(^{#1}\)\fi}
	\caption{First stage: Propensity to receive \textit{Avancemos}}
	\label{tab:instruments}
	\begin{threeparttable}
		\begin{tabular}{l*{2}{c}}
			\hline\hline
			&\multicolumn{1}{c}{Daughters}&\multicolumn{1}{c}{Sons}\\
			&\multicolumn{1}{c}{} &\multicolumn{1}{c}{} \\
			\hline
			Percentage of households with \textit{Avancemos}  
			&      
			2.138\sym{***}&      
			1.449\sym{***}\\
			&     (0.271)         &     (0.256)         \\
			\hline
			Controls &     Yes       
			&     Yes        
			\\
			Year and geographical effects&   
			Yes        &     Yes       
			\\
			Observations        &        1,090         &       
			1,357         \\
			\(R^{2}\)           &         0.146            &      0.127              
			\\
			\hline\hline	
		\end{tabular}
		\begin{tablenotes}\footnotesize
			\item[] \footnotesize Standard errors between parentheses. \sym{*} \(p<0.05\), \sym{**} \(p<0.01\), \sym{***} \(p<0.001\).
			\item[] \footnotesize Control variables include the teens' age, parents' education, father's occupation, work industry, insurance, and the number of younger children in the household. The complete table is in Appendix \ref{app:mte_results}.
		\end{tablenotes}
	\end{threeparttable}
\end{table}

The exogeneity of the instrument for condition (ii) is not obvious. Exogeneity should hold if the stable unit treatment value assumption (SUTVA) for the outcomes and the instrument's design, which takes spillovers in treatment uptake into account, are true. I control for a set of individual and household characteristics, including age, parental education, a marker for rural residence and region, and the father's job characteristics, such as occupation and industry, to strengthen the exogeneity. However, there can be concerns related to the endogeneity of where the household decides to leave. First, households are not located randomly. This endogeneity complements the idea of the instrument by taking into account the peer effects of the neighborhood in the treatment take-up. Secondly, I might be confusing the effect of \textit{Avancemos} on labor outputs with the effect of infrastructure on labor outputs because neighborhoods with a lot of high schools can have a lot of other types of infrastructure. Even though this can bias my results, a more developed neighborhood has more private high schools that do not give the option to apply for \textit{Avancemos}, hence moving in the same direction as my instrument.

\subsubsection{Results}
\begin{figure}[h!]
	\centering
	\caption{\textit{Avancemos} on teen's schooling decision: MTE.}%
	\label{fig:mte_school}%
	\begin{subfigure}{.35\linewidth}
		\centering\includegraphics[width=1.3\linewidth]{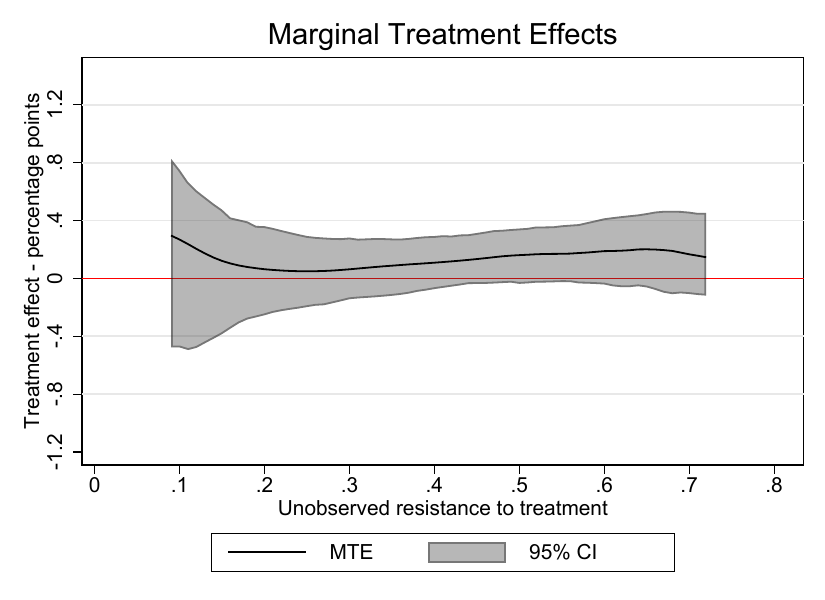}
		\caption{Daughters.}
	\end{subfigure}\hspace{18mm}
	\begin{subfigure}{.35\linewidth}
		\centering\includegraphics[width=1.3\linewidth]{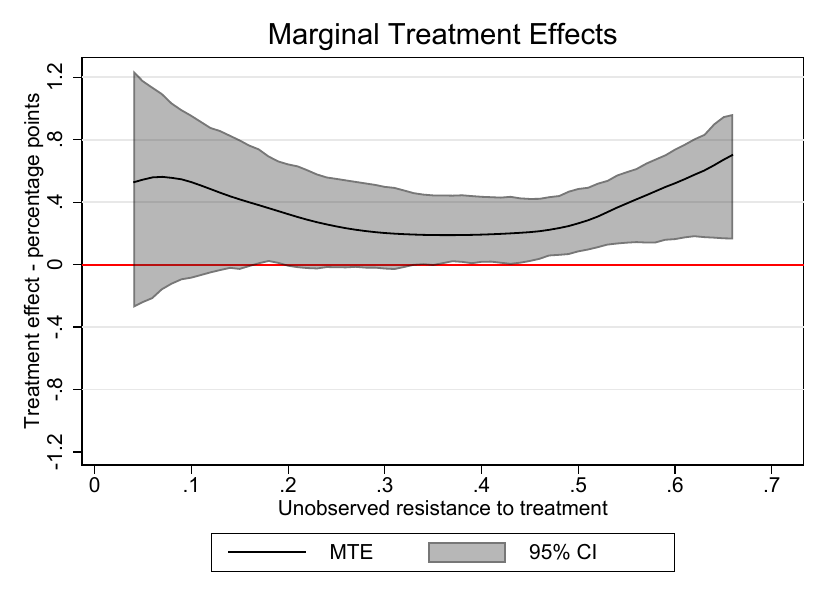}
		\caption{Sons.}
	\end{subfigure}
\end{figure}

I estimate the MTE separately for the daughters and sons. To estimate the MTE for all quantiles, there should be enough units of treated and nontreated households for each value of the propensity score. However, Figure \ref{fig:pscores} shows that common support is not satisfied in my sample. I estimate the MTE in the daughters' sample up to the 69th percentile, and in the sons' sample up to the 65th percentile. My outcome variables are the teens' schooling decision, the fathers' labor supply, the mothers' employment status, and the domestic supply for the three members. Figure \ref{fig:mte_school} shows the results for the unobservable part of the treatment effect for the teen's schooling decision\footnote{Note: MTEs are calculated with the separate approach by \cite{heckman2005structural} using a polynomial of degree 1 and a semiparametric local polynomial of degree 2. Standard errors are computed with a bootstrap procedure (250 replications).}. The MTE has an upward-sloping shape for higher values of unobserved treatment resistance, indicating a pattern of reverse selection on gains. Whereas sons in households with a low resistance to \textit{Avancemos} do not have a statistically significant effect on their attendance in high school, sons in households with higher resistance have a positive and significant effect. There is no significant effect on daughters. I show the estimation results of the observable part of the treatment effect in Appendix \ref{app:mte_results}. The main differences in outcomes come from teens' age, mother's education, and father's occupation.

\begin{figure}[h!]
	\centering
	\caption{\textit{Avancemos} on mother's employment status: MTE.}%
	\label{fig:part_mother}%
	\begin{subfigure}{.35\linewidth}
		\centering\includegraphics[width=1.3\linewidth]{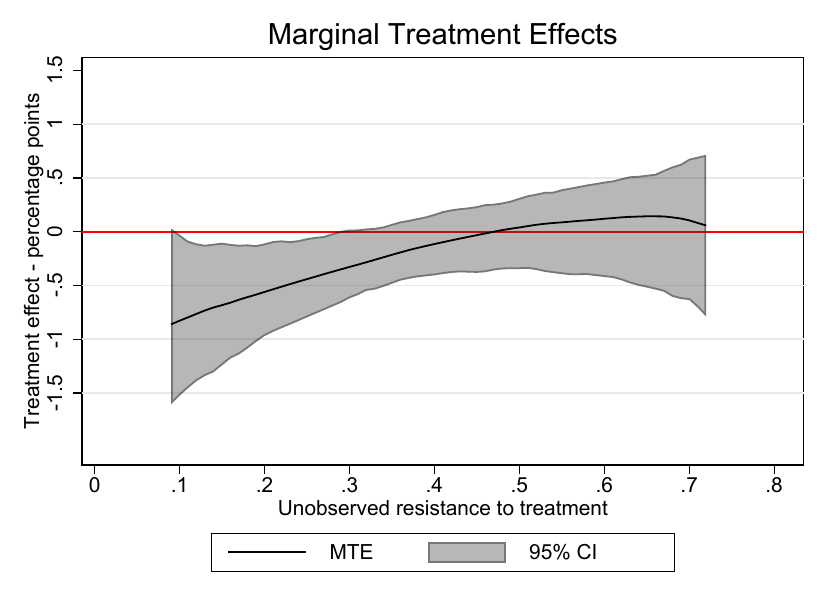}
		\caption{Daughters.}
	\end{subfigure}\hspace{18mm}
	\begin{subfigure}{.35\linewidth}
		\centering\includegraphics[width=1.3\linewidth]{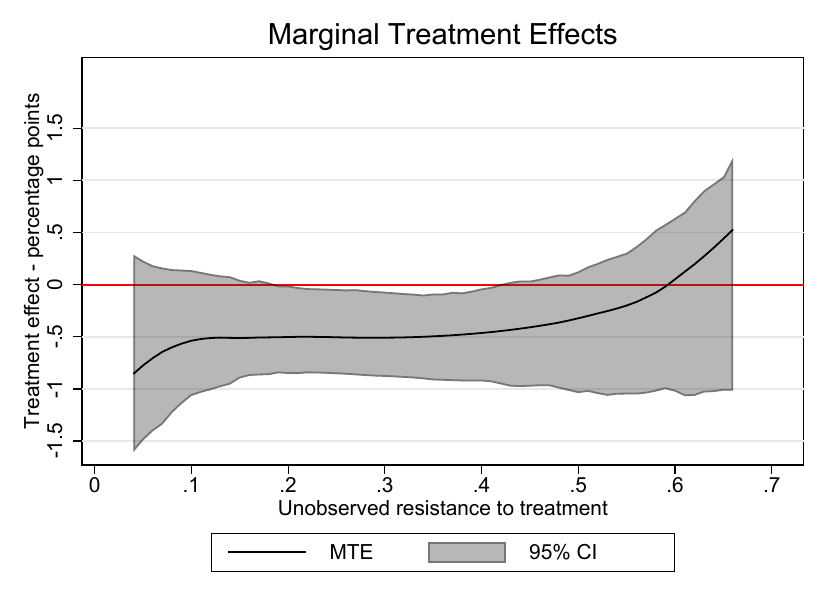}
		\caption{Sons.}
	\end{subfigure}
\end{figure}

\begin{figure}[h!]
	\centering
	\caption{\textit{Avancemos} on father's labor supply: MTE.}%
	\label{fig:labour_father}%
	\begin{subfigure}{.35\linewidth}
		\centering\includegraphics[width=1.3\linewidth]{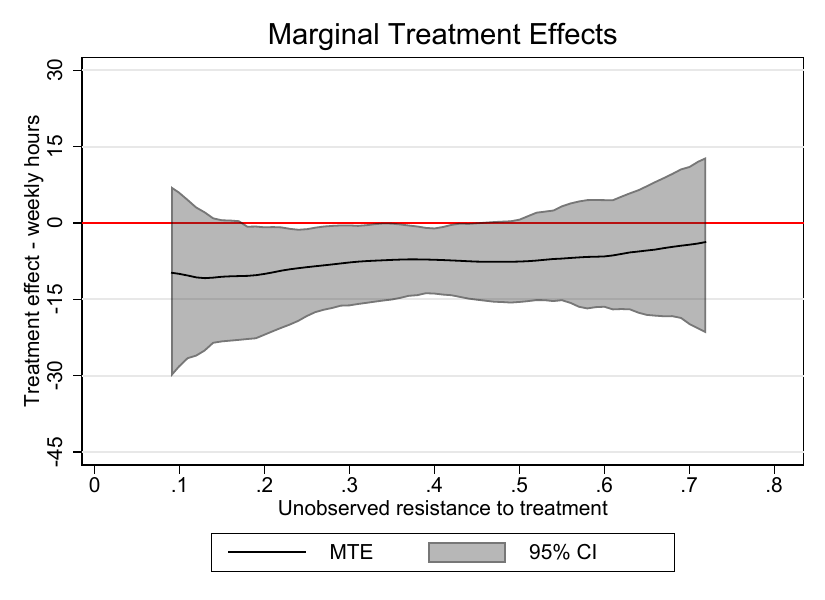}
		\caption{Daughters.}
	\end{subfigure}\hspace{18mm}
	\begin{subfigure}{.35\linewidth}
		\centering\includegraphics[width=1.3\linewidth]{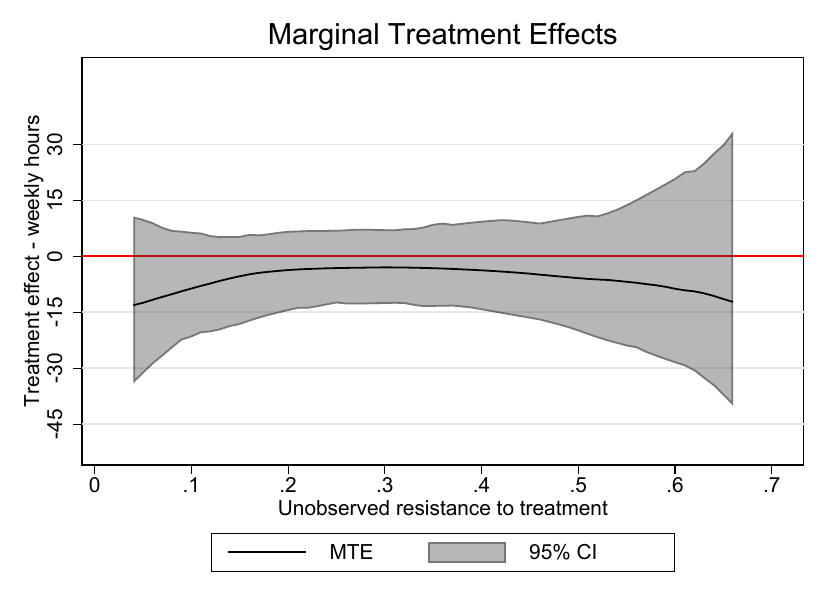}
		\caption{Sons.}
	\end{subfigure}
\end{figure}

Figure \ref{fig:part_mother} shows a negative and significant effect on the mother's employment status, regardless of whether she has a daughter or a son. Although the effects vary depending on the unobserved resistance, mothers in households with a higher resistance benefit most from \textit{Avancemos}. Furthermore, the slopes are pretty different, indicating a negative selection for households with daughters but no selection on gains if the teen is a son (flat MTE). Figure \ref{fig:labour_father} shows the results of the unobserved part of the MTE for the father's labor supply. For the fathers, there is no gender difference selection into gains, but fathers with a daughter benefit from \textit{Avancemos} by decreasing their labor supply. Finally, none of the three household members experiences any significant changes in domestic supply. The graphs and the rest of the results are in Appendix \ref{app:mte_results}.

To summarize, \textit{Avancemos} reallocates household time. The daughter continues to attend high school, her mother is less likely to work, and her father reduces his labor supply. A standard income effect can explain these changes, in which \textit{Avancemos}, as extra income, allows the parents to enjoy more leisure. However, it is not the same if the teen is a son. In that case, he is more likely to attend high school, and his mother is less likely to work in the labor market, but his father's labor supply remains unchanged. One explanation could be that the son bargains cooperatively with his parents and receives a share of the household's resources. If this is the case, the extra income from \textit{Avancemos} is shared differently between parents and sons. To test whether the son is a decision-maker and to quantify his bargaining effect, I present a collective household model in the section below.

\section{Theoretical model}
The common models of household behavior are the unitary household model, which assumes that households behave as one single decision-maker, and the collective household model, which considers households as a group of individuals, each with their utility. In a collective household model, members allocate resources according to their bargaining power, which depends on their outside options as in a cooperative game.

The existence of cooperative bargaining in households is documented in the literature, but it excludes the possibility of teens bargaining with their parents about time allocation. Children are typically regarded as public goods in the utility of their parents. My data, however, show behavior that does not fit the traditional father and mother collective model. In this section, I present a collective model with parents and teens as decision-makers and a unitary model in a time-use framework. The models allow me to test the income pooling-hypothesis of the unitary model and the bargaining constraints obtained from the collective model.

The models include parents' labor supply, teens' schooling decisions, and the production of a domestic public good. Let $i=p,t$ denote the parents\footnote{I combine parents as a single decision-maker for two reasons. First and foremost, my goal is the bargaining between parents and teens, not about disentangling fathers and mothers. Second, data constraints prevent me to consider a collective model with three decision-makers.} and the teen. The household produces a public good through the production function $f^{K}(h^{p},h^{t})$, which takes domestic work $(h^{p},h^{t})$ as input. This function is twice continuously differentiable in all its arguments, strictly increasing and strongly concave. The good created from this production function can be interpreted as the satisfaction of a clean and warm house. The household faces the same restrictions in the unitary and collective models. I present these restrictions first. For the parents, their time constraint\footnote{Time is normalized to 1.} is between leisure, $l^{p}$; labor market work, $m^{p}$ with the father always working, and domestic work, $h^{p}$:
\begin{equation*}
	l^{p} + m^{p} + h^{p} = 1.
\end{equation*}
The teen allocates her time between school ($s^{t}=1$), work ($s^{t}=0$) and domestic work $h^{t}$.
\begin{equation*}
	l^{t}  + h^{t} + S\mathbf{1}\{s^{t}=1\} + \overline{m}^{t}\mathbf{1}\{s^{t}=0\}  = 1
\end{equation*}
where $S$ is the amount of time spent in school and $\overline{m}^{t}$ her working time. Each member consumes a Hicksian composite good $C$ (= $C^{p} + C^{t}$). The price of the consumption good is set to 1. The household acquires goods according to the following budget constraint:
\begin{equation*}
	C =  w^{p}m^{p} + w^{t}\overline{m}^{t}\mathbf{1}\{s^{t}=0\} + y + y_{A}\mathbf{1}\{Avancemos=1\}\mathbf{1}\{s^{t}=1\}
\end{equation*}
where $w^p$ is the parents' hourly wage, $w^t$ is the teen's hourly wage if she works, $y$ is nonlabor income and $y_{A}$ is \textit{Avancemos}' transfer.

For the unitary model, the household behaves as a single unit according to a twice continuously differentiable, strictly monotonic, and strongly concave utility function $U^{H}(l^{p},l^{t},C,f(h^{p},h^{t}))$. It maximizes the following program:
\begin{subequations}\label{model/unitary}
	\begin{equation}\tag{\ref{model/unitary}}
		\underset{m^{p},s^{t},h^{p},h^{t},C}{\max} \; \; U^{H}(l^{p},l^{t},C,f^{K}(h^{p},h^{t}))
	\end{equation}
	\begin{numcases}{s.t.}
		C =  w^{p}m^{p} + w^{t}\overline{m}^{t}\mathbf{1}\{s^{t}=0\} + y + y_{A}\mathbf{1}\{Avancemos=1\}\mathbf{1}\{s^{t}=1\} \\
		0 < m^{p}\leq 1, \; \; s^{t} \in \{0,1\}, \; \; m^{t} \in \{0,\overline{m}\} \\
		l^{p} + m^{p} + h^{p} = 1 \\
		l^{t} + h^{t} + \overline{m}^{t}\mathbf{1}\{s^{t}=0\} + S\mathbf{1}\{s^{t}=1\}  = 1 
	\end{numcases}
\end{subequations}

Following \citet{chiappori1992collective}, I assume in the collective model that each member has an egoistic, twice continuously differentiable, strictly monotonic, and strongly concave utility function $U^{i}(l^{i},C,f^{K}(h^{p},h^{t}))$. The members' bargaining function depends on the parents' labor income $w^{p}$, the teen's wage $w^{t}$, and the household's nonlabor income $y$. As is usual in the literature, I assume that the members choose Pareto efficient intrahousehold allocations
\footnote{A simple argument in favor of this assumption is that members are aware of each other's preferences and are unlikely to disregard Pareto-improving decisions as a result of their interaction. For more information about the validation of Pareto efficiency, see the surveys \citet{vermeulen2002collective} and \citet{chiappori2017static}.}. The household's maximization program is:
\begin{subequations}\label{model/hh}
	\begin{equation}\tag{\ref{model/hh}}
		\underset{m^{p}, h^{p}, C^{p}, m^{t}, h^{t}, C^{t}}{\max} \; \; \lambda^{p}(w^{p},w^{t},y) U^{p}(l^{p},C^{p},f^{K}(h^{p},h^{t})) +
		\lambda^{t}(w^{p},w^{t},y) U^{t}(l^{t},C^{t},f^{K}(h^{p},h^{t}))
	\end{equation}
	\begin{numcases}{s.t.}
		\lambda^{p}(w^{p},w^{t},y) +    \lambda^{t}(w^{p},w^{t},y) = 1 \\
		C^{p} + C^{t} =  w^{p}m^{p} + w^{t}\overline{m}^{t}\mathbf{1}\{s^{t}=0\} + y + y_{A}\mathit{Pr}(Avancemos=1)\mathbf{1}\{s^{t}=1\}  \label{b}\\
		0 < m^{p}\leq 1, \; \; s^{t} \in \{0,1\}, \; \; m^{t} \in \{0,\overline{m}\} \\
		l^{p} + m^{p} + h^{p} = 1 \\
		l^{t} + h^{t} + \overline{m}^{t}\mathbf{1}\{s^{t}=0\} + S\mathbf{1}\{s^{t}=1\}  = 1 
	\end{numcases}
\end{subequations}
where $\lambda^{i}(w^{p},w^{t},y)$ is the Pareto weight for the parents and the teen, which has a direct relationship to their bargaining power.

A two-stage model can be used to represent both the unitary and collective models. The first stage of the unitary model determines how much public good is produced. In the second stage, the household maximizes utility with residual income, conditional on the quantity of public goods produced. The first stage in the collective model is similar; the household agrees to produce a certain level of public good $\bar{f}^{K}$. However, conditional on the public good created and cooperative bargaining, parents and teens receive a share of the residual income. This distribution takes place at the Pareto frontier of household decision-making and is represented with the conditional sharing rule $\rho^{i}(w^{p},w^{t},y | \bar{f}^{K}), \; i=\{p,t\}$. In the second stage, each member decides freely on their level of leisure and consumption with their conditional share of income.

It is important to emphasize two points about these models. First, public good production is efficient in both cases, and the Bowen-Lindahl-Samuelson\footnote{This condition states that at the optimal level of the public good, the sum of the parents and teen's marginal benefit from the public good is equal to its marginal cost.} condition holds \citep{blundell2005collective}. Second, the difference in household behavior is reflected in both the teen's school attendance and the labor supply of the parents. Income is pooled in the unitary case, such that an increase in nonlabor income or in the teen's wage has the same effect on the parents' labor supply (if the teen only works). The extra income has a standard income effect on the teen's schooling decision. In the collective model, changes in a household's income include the bargaining effect. Because the teen's wage increases the teen's bargaining power, there is a difference in the parents' labor supply depending on whether extra income comes from nonlabor income or an increase in the teen's wage. There is also an effect even if the teen attends school and does not have a wage because her potential wage gives her bargaining power. The bargaining effect is reflected in the reservation wage that defines the teen's schooling and work decisions. Changes in her parents' wages have an impact on the reservation wage due to changes in the bargaining process, and thus her share of resources. The mathematical representation of these effects is in Appendix \ref{app:restrict_unitary} for the unitary model and in Appendix \ref{app:restrict_collective} for the collective model.

\subsection{Identification}
The distribution of wages and nonlabor income are the primary sources of identification in the models. They are sufficient to identify all of the parameters in the unitary case. In the collective model, however, identifying the sharing rule is more difficult. I use two results from the literature to identify the sharing rule. First, \citet{cherchye2012married} provide the identification of the home production function within a collective model. The issue is that, in general, the variation in wages and nonlabor income also impacts domestic production. A solution is to use "production shifters", variables that impact the home production function but not preferences. These variables enable estimating the substitution parameter in a CES function, fixing the level of public goods produced in the household. Once domestic production is fixed, the wage and nonlabor variation identify the bargaining function \citep{chiappori1992collective}. The children's ages and house characteristics are frequently used as production shifters in literature.  I am unable to use these variables because I condition my sample based on the teen and younger children's ages. I currently impose Cobb Douglass production function with a substitution parameter equal to one. It allows me to identify the model without using production shifters.

The second identification result is provided by \citet{blundell2007collective} for collective models in the case in which one of the outputs is on the extensive margin, the teen's schooling decision in my case. To recover the sharing function from the estimation, it is necessary to make a double indifference assumption. This assumption states that parents and teens are indifferent about going to school across the schooling frontier. This assumption allows for the sharing rule to be continuous across the frontier, keeping the Pareto assumption across it. Finally, the conditional sharing rule is identified up to a constant. Because I do not observe individual private consumption or the output of the household production function, it is impossible to distinguish between household heterogeneity in outputs and sharing rules \citep{chiappori1992collective}.

Both results provide nonparametric identification. Following the literature, I use a parametric specification to estimate the model.

\section{Parametric specification}
In this section, I describe the parametric model that allows me to estimate and test the unitary and collective models. Using the models' two-stage framework, I derive the equations I estimate and the constraints I test. I start with the estimation of the public good production, and then parental labor supply and teen schooling. 	

For the public good production, I assume it is produced with Cobb Douglas technology with a productivity parameter $\alpha$:
\begin{equation*}
	f^{K}(h^p,h^t) =  (h^p)^{\alpha} \;  \; (h^t)^{1-\alpha}
\end{equation*}
and has a price equal to:
\begin{equation*}
	g^{K}(w^{p},w^{t}) = \bigg( \frac{w^{p}}{\alpha} \bigg)^{\alpha} \;  \; 
	\bigg( \frac{w^{t}}{1-\alpha} \bigg)^{1-\alpha}.
\end{equation*}
If households produce effectively while minimizing costs, regardless of their behavior, the amount produced satisfies the Bowen-Lindahl-Samuelson condition of efficient public good provision. Under this assumption and applying Shepard's lemma, the domestic supply Hicksian demands are:
\begin{align*}
	h^{p}(w^{p},w^{t}) &=\bigg( \frac{w^{p}}{\alpha} \bigg)^{\alpha-1} \;  \; 
	\bigg( \frac{w^{t}}{1-\alpha} \bigg)^{1-\alpha} \;  \; \overline{f}^{K}  \\
	h^{t}(w^{p},w^{t}) &=\bigg( \frac{w^{p}}{\alpha} \bigg)^{\alpha} \;  \; 
	\bigg( \frac{w^{t}}{1-\alpha} \bigg)^{-\alpha} \;  \; \overline{f}^{K}.
\end{align*}
Taking the differences in log, I obtain the following equation:
\begin{equation}\label{eq:home_prod}
	\textrm{ln}\bigg( \frac{h^{p}}{h^{t}} \bigg) (w^{p},w^{t}) = \textrm{ln}\bigg( \frac{w^{t}}{w^{p}} \bigg) + \textrm{ln}\bigg( \frac{\alpha}{1- \alpha} \bigg) + u_{h},
\end{equation}
where $u_{h}$ denotes a measurement error. This equation estimates the productivity parameter $\alpha$.

\subsection{Unitary model}
For the unitary model, I assume a semilog indirect utility function for the household defined as:
\begin{equation}\label{eq:unitary_utility}
	v^{H}(w^p,y^*,\overline{f}^{k}) = \frac{\textrm{exp}(\theta^{y}w^{p})}{\theta^{y}} \bigg(\theta^{p}_{w}\textrm{ln}w^{p} + \theta^{y}y^{*} + \theta^{H}_{K} \textrm{ln}\overline{f}^{K} \bigg) - \frac{\theta^{p}_{w}}{\theta^{y}} \int_{-\infty}^{\theta^{y}w^{p}} \frac{\textrm{exp}(t)}{t} dt,
\end{equation}
where $y^{*} = y + w^{t}$ if the teen works or  $y^{*} = y + y_{A}\mathbf{1}\{Avancemos\}$ if the teen goes to school. The Bowen-Lindahl-Samuelson condition in this model is:
\begin{equation*}
	g^{K}(w^{p},w^{t}) = \frac{1}{\overline{f}^{K} } \frac{\theta^{H}_{K}}{\theta^{y}}
\end{equation*}
For the second-stage model, the Marshallian labor supply for the parents follows Roy's identity in equation (\ref{eq:unitary_utility}):
\begin{equation*}
	M^{p}(w^{p},w^{t},y,\overline{f}^{K}) = \theta^{p}_{w}\textrm{ln}w^{p} + \theta^{y}y^{*} + \theta^{H}_{K} \textrm{ln}\overline{f}^{K}.
\end{equation*}
I replace $\overline{f}^{K} $ with the Bowen-Lindahl-Samuelson condition and obtain:
\begin{equation}\label{eq:marsh_uni}
	M^{p}(w^{p},w^{t},y) = \theta^{p}_{w}\textrm{ln}w^{p} + \theta^{y}y^{*} + \theta^{H}_{K} \big[ \alpha \textrm{ln}w^{p}  + (1-\alpha)\textrm{ln}w^{t} - \alpha \textrm{ln}\alpha - (1-\alpha)\textrm{ln}(1-\alpha) \big].
\end{equation}
This equation is estimated using a switching regression model based on the teen's educational status:
\begin{equation}
	\begin{split}
		m^{p}(w^{p},w^{t},y) &= A^{*}_{p}\textrm{ln}w^{p} + A_{t}w^{t} + A_{y}y + \delta\textrm{ln}w^{t} + \textbf{X}_{w}\beta_{w}  + u_{w}, \; \; \textrm{if} \; s^{t}=0 \\
		m^{p}(w^{p},w^{t},y) &= a^{*}_{p}\textrm{ln}w^{p} + a_{t}w^{t} + a_{y}y +  \delta\textrm{ln}w^{t} + \textbf{X}_{s}\beta_{s}  + u_{s}, \; \; \textrm{if} \; s^{t}=1
	\end{split}
\end{equation}
where $\textbf{X}$ is a vector of covariates and $u_{w}$ and $u_s$ are measurement errors. For the teen's schooling decision, I use a latent index:
\begin{equation}\label{eq:schooling_uni}
	\begin{split}
		(s^t)^{*} &= b + b_{t} w^{t} + b_{p} \textrm{ln} w^{p} +  b^{t}_{y} y + \textbf{X}_{t}\beta_{t} + u_{t} \\
		s^t &= 1 \; \; \textrm{if} \; \Phi((s^t)^{*}\geq 0), \; \;   s^t = 0 \; \; \textrm{otherwise}
	\end{split}
\end{equation}
$u_t$ represents a measurement error. There are two key points to note in this equation. First, because it represents the schooling decision frontier, the public good is not included. Second, it is possible to recover the reservation wage when $s^t = 0$, where the frontier parameters are:
\begin{equation}\label{eq:reservation}
	\gamma_p = -\frac{b_{p}}{b_{t}}, \; \; \; \gamma_y = -\frac{b_{y}}{b_{t}}.
\end{equation}

The novelty of my model is the introduction of home production in the estimation. It establishes a direct relationship between the reduced-form parameters and the Cobb-Douglas technology. Explicitly, in the estimation, $\delta=\theta^{H}_{K}\;(1-\alpha)$ and $A^{*}_{p} = A_{p} + \theta^{H}_{K}\;\alpha$, where $A_{p}$ has a direct relationship to equation (\ref{eq:marsh_uni}). Similarly, for $a^{*}_{p}$. Once the technology parameters are recovered, the remaining parameters are used to test the unitary model restrictions \citep{blundell2007collective}:
\noindent
\begin{multicols}{2}
	\noindent
	\begin{equation}\label{rest_u1}\tag{U1}
		\begin{split}
			A_{t} &= A_{y} \\
			a_{t} &= 0
		\end{split}
	\end{equation}
	\columnbreak
	\begin{equation}\label{rest_u2}\tag{U2}
		\begin{split}
			(1+\gamma_{y})&(a_{y}-A_{y}) = 0 \\
			A_{y}\gamma_{p} &= (1+\gamma_{y})a^{*}_{p} - A^{*}_{p}
		\end{split}
	\end{equation}
\end{multicols}

The unitary model can be tested using these two sets of constraints.  The null hypothesis for these restrictions is that the household acts as a single unit decision-maker. Under the null, restrictions \ref{rest_u1} refer to the household income distribution and how it affects the labor supply of the parents. There is the same effect of an increase in the teen's wage and non-labor income on the parents' labor supply. If the teen does not work, there is no effect on her potential wage. Restrictions \ref{rest_u2} refer to income and substitution effects caused by shifts in the labor supply of the teen's parents on her decision to enroll in school. Increases in non-labor income result in higher reservation wages, allowing the teen to continue their education for longer. The same holds for a raise in the parents' salary.

\subsection{Collective model}
The cooperative resource bargaining between parents and teenagers is the collective model's key characteristic. This negotiation establishes Pareto weights, which are converted into a unique resource share ($\rho^i$) for each household decision-maker. I assume that the parents' preference for market work and consumption has the following indirect utility belonging to the semilog labor supply equations \citep{stern1986specification}:
\begin{equation}
	v^{p}(w^p,\rho^{p},\overline{f}^{k}) = \frac{\textrm{exp}(\theta^{p}_{\rho}w^{p})}{\theta^{p}_{\rho}} \bigg(\theta^{p}_{w}\textrm{ln}w^{p} + \theta^{p}_{\rho}\rho^{p} + \theta^{p}_{K} \textrm{ln}\overline{f}^{K} \bigg) - \frac{\theta^{p}_{w}}{\theta^{p}_{\rho}} \int_{-\infty}^{\theta^{p}_{\rho}w^{p}} \frac{\textrm{exp}(t)}{t} dt.
\end{equation}
The teen's indirect utility is:
\begin{equation}
	v^{t}(w^t,\rho^{t},\overline{f}^{k}) = \theta^{t}_{\rho}\rho^{t} + \theta^{t}_{K} \textrm{ln}\overline{f}^{K}.
\end{equation}
These utilities define the production of the public good and the conditional distribution of resources \citep{blundell2005collective}. Under efficient production, the Bowen-Lindahl-Samuelson condition is:
\begin{equation*}
	g^{K}(w^{p},w^{t}) = \frac{1}{\overline{f}^{K} } \bigg(\frac{\theta^{p}_{K}}{\theta^{p}_{\rho}} + \frac{\theta^{t}_{K}}{\theta^{t}_{\rho}}\bigg).
\end{equation*}
I assume the conditional bargaining function has the form
\begin{equation}\label{eq:sharing}
	\rho(w^{p},w^{t},y| \bar{f}^{K}) = \psi_{p}\textrm{ln}w^{p} + \psi_{t}w^{t} + \psi_{y} y,
\end{equation}
where I use the level of teen earnings rather than the log because it allows me to nest the income pooling hypothesis with nonlabor income, the use of the log in the parents' wage is to connect it to the parents' labor supply. I assume for the second stage problem that $\rho(w^{p},w^{t},y| \bar{f}^{K}) = \rho^{t}$, which implies $\rho^{p} = y^* - \rho^{t} - g^{K}(w^p,w^p)\overline{f}^{K}$. Applying Roy's identity to the parents' indirect utility function gives their (conditional) Marshallian labor supply:
\begin{equation*}\label{eq:marsh_collective}
	M^{p}(w^{p},\rho^{p},\overline{f}^{K}) = \theta^{p}_{w}\textrm{ln}w^{p} + \theta^{p}_{\rho}\rho^{p} + \theta^{p}_{K} \textrm{ln}\overline{f}^{K}
\end{equation*}
Substituting the sharing rule and including the Bowen-Lindahl-Samuelson condition gives:
\begin{equation*}
	M^{p}(w^{p},w^{t},y^*) = (\theta^{p}_{w}-\theta^{p}_{\rho}\psi_{y} + \theta^{p}_{K}\alpha)\textrm{ln}w^{p} +  (\theta^{p}_{\rho}(1-\psi_{t}))w^{t} +(\theta^{p}_{\rho}(1-\psi_{y}))y + \theta^{p}_{K}(1-\alpha)\textrm{ln}w^{t} + U
\end{equation*}
where $U = \theta^{p}_{K} \textrm{ln}\bigg(\frac{\theta^{p}_{K}}{\theta^{p}_{\rho}} + \frac{\theta^{t}_{K}}{\theta^{t}_{\rho}}\bigg) - \theta^{p}_{\rho}\bigg(\frac{\theta^{p}_{K}}{\theta^{p}_{\rho}} + \frac{\theta^{t}_{k}}{\theta^{t}_{\rho}}\bigg) + \theta^{p}_{K}(-\alpha \textrm{ln}\alpha - (1-\alpha)\textrm{ln}(1-\alpha))$. As in the case of the unitary model, I estimate this equation through a switching regression regime defined by the teen's schooling status:
\begin{equation}
	\begin{split}
		m^{p}(w^{p},w^{t},y) &= A^{*}_{p}\textrm{ln}w^{p} + A_{t}w^{t} + A_{y}y + \delta\textrm{ln}w^{t} + \textbf{X}_{w}\beta_{w}  + u_{w}, \; \; \textrm{if} \; s^{t}=0 \\
		m^{p}(w^{p},w^{t},y) &= a^{*}_{p}\textrm{ln}w^{p} + a_{t}w^{t} + a_{y}y^* +  \delta\textrm{ln}w^{t} + \textbf{X}_{s}\beta_{s}  + u_{s}, \; \; \textrm{if} \; s^{t}=1
	\end{split}
\end{equation}
Finally, the teen's schooling decision is modeled as in the unitary case with equation (\ref{eq:schooling_uni}). It is important to highlight that these reduced-form equations are the same as in the unitary model. However, depending on which model we accept or reject, the estimated parameters have different interpretations. Again, the novelty here is the introduction of home production. Using direct mapping to recover Cobb Douglas technology, the rest of the parameters are used to test the restrictions of the collective model, as in \cite{blundell2007collective}:
\begin{equation}\label{rest_c1}\tag{C1}
	\frac{A_{t} - a_{t}}{A_{y}- a_{y}} =             -\frac{1}{\gamma_{y}}, \; \; \; \frac{A^{*}_{p} - a^{*}_{p}}{A_{y}- a_{y}} =     \frac{\gamma_{p}}{\gamma_{y}}
\end{equation} 

Restrictions \ref{rest_c1} allow us to test the collective model. The household behaves as though there are two decision-makers, the parents and the teen, according to the collective model's null hypothesis. In contrast to the unitary model, households in a collective framework consider how non-labor income and wages may alter the household's budget constraint and the bargaining power of the decision-maker who brings them. Restrictions \ref{rest_c1} are a simplified version of the restrictions in the parents' labor supply and the teen's schooling decision.

\subsection{Stochastic specification}

The estimation consists of four equations: the home production function, the teen's schooling participation, and the parents' switching labor supply. To allow for unobserved factors within households, I follow \citet{bonnal1997evaluating} and assume that the four measurement errors ($u_{h},u_{t},u_{w},u_{s}$) are generated by a common normally distributed random variable $\eta$ such as:
\begin{equation*}
	u_{i,j} = \eta_{i} \chi_{j} + \varepsilon_{i,j}
\end{equation*}
where $i$ is a household, $j$ is one of the equations and $\varepsilon_{i,j}$ is an i.i.d. shock. 

An important part of the paper is \textit{Avancemos}, which is endogenous in this estimation for two different reasons. First, the households who benefit from it must send their teen to school because it is the condition of the transfer. Because of this, I only take into account the households that do not receive the transfer when I estimate the participation in schooling. Second, as \textit{Avancemos} affects the amount of nonlabor income in the household, using this variable generates an omitted endogeneity bias due to the selection into treatment. I employ the control function method suggested by \citet{blundell2003endogeneity} to correct this bias. I use the amount of \textit{Avancemos} and household covariates to regress nonlabor income. I then use the residuals as a control variable in the parents' labor supply equations and the teen's choice of school. With the residuals serving as a covariate to correct the inconsistency of my estimates, much like a Heckman two-step correction, I treat endogeneity as an omitted variable problem in this way.  Appendix \ref{app:control_function} contains the regression results in detail.

Last, I impute the teen's wages to estimate the model for those households in which the teen goes to school. For the imputation, I use a larger subsample from the ENAHO because there are not enough observations with wages in my sample, especially for daughters. It consists of individuals considered children in the household, aged 15 to 25, without a high school diploma. It differs from my sample in that it covers a variety of households, such as single parents and extended households. I estimate wages individually for sons and daughters. I explain this imputation in detail in Appendix \ref{app:imputation}. 

I use maximum likelihood to estimate the unrestricted model, the unitary model, and the collective model. The likelihood contribution for household $i$ is:
\begin{equation*}
	\begin{split}
		L_{i} = \big[\Pr((s^t)^{*}\geq0) \times \mathit{f}(m_{i}^{p}|(s^t)^{*}\geq0)\big]^{\mathbf{1}\{s_{i}^{t}=1 \}} &\times \big[\Pr((s^t)^{*}<0) \times 
		\mathit{f}(m_{i}^{p}|(s^t)^{*}<0) \big]^{\mathbf{1}\{s_{i}^{t}=0 \}} \\
		&\times  \mathit{f}(\textrm{ln}(h_{i}^{p}/h_{i}^{t}))
	\end{split}
\end{equation*}

\section{Results}
The main result is determining whether the teen is a decision-maker. For this, I use the likelihood ratio test for the unitary restrictions \ref{rest_u1} and \ref{rest_u2} and the collective restrictions \ref{rest_c1}. The null hypothesis for the unitary model is that the household acts as a single unit decision-maker. The null hypothesis for the collective model is that the household behaves with two decision-makers: the parents and the teen. The alternative hypothesis is not clear for both tests because it is difficult to tell from which channel the rejection of the null hypothesis comes. The estimation results are presented in Appendix \ref{app:estimates}. Starting with the son, the likelihood ratio statistic for the unitary model restrictions is 26.73 with a p-value of 0.00003; hence, I reject that a household with a son behaves as the unitary model predicts. For the collective model restrictions, the likelihood ratio statistic is 1.28 with a p-value of 0.53; hence, I do not reject that the son is a decision-maker. For the daughter\footnote{I estimate the domestic outputs for sons both jointly and separately with the other outcomes, and I obtain equivalent results. I assume that the same result would be valid for daughters. I do this because I impute most of the wages for the daughters, concentrating variation around the mean.}, the likelihood ratio statistic for the unitary model is 21.98 with a p-value of 0.002; hence, I reject that a household with a daughter behaves as the unitary model predicts. For the collective model, the likelihood ratio statistic is 6.93 with a p-value of 0.03; hence, I also reject the daughter as a decision-maker. 

The difference between daughters and sons being decision-makers or not is reflected in their opportunity costs. Going to school suggests that the son has less negotiating power with his parents, which lowers his income share. The opportunity cost for daughters is at the household level and not on her own because the daughter lacks bargaining power and attends school as part of the decision made by the household. This opportunity cost is in the model via a reservation wage. Recovering the reservation wages for sons and daughters is possible with equation \ref{eq:reservation}:

\begin{align*}
	w^{r}_{son} &= \kappa_{son} + \underset{(0.070)}{0.405} \; \textrm{ln}w_{p} + \underset{(0.158)}{0.493} \;  y, \; \; \textrm{if teen is a son} \\
	w^{r}_{daughter} &= \kappa_{daughter} + \underset{(0.410)}{0.359} \; \textrm{ln}w_{p} + \underset{(0.430)}{0.668} \;  y, \; \; \textrm{if teen is a daughter.} \\
\end{align*}
The coefficients show that daughters' reservation wages increase more than sons' for additional non-labor income and the parents' wages, even though I cannot test the difference between them (because they were estimated using different samples).

My results show that daughters are not included in the decision-making process regarding their potential labor-force participation. Most daughters seem to be led to prioritizing their education over work. This could be because work opportunities are less appealing - lower wages, possibly fewer job offers. On the other hand, the decision to keep daughters in school may be motivated by the high cost of preceding education to support the household. Indeed, schooling gives daughters a lot of time to participate in housework, whereas a job would make them less available. Furthermore, the conservative Costa Rican society may stigmatize young women who work, which may limit their opportunities. Further research is needed to disentangle these potential mechanisms.

Because the son is not rejected as a decision-maker, I can recover the bargaining function by mapping the estimated and structural parameters. The mapping is explained in detail in Appendix \ref{app:recovering}. The son's bargaining function from equation \ref{eq:sharing} is\footnote{Standard errors are estimated using the Delta Method.}:
\begin{equation*}\label{eq:sharingfunct}
	\begin{array}{ll}
		\rho_{t} &= \kappa_{1} + \underset{(0.223)}{1.146} \; w_{t} + \underset{(0.753)}{1.796} \; \textrm{ln}w_{p} + \underset{(0.505)}{2.190} \;  y, \; \; \textrm{if} \; s^{t}=0 \\
		\rho_{t} &= \kappa_{0} + \underset{(0.051)}{0.821}\bigg(\underset{(0.223)}{1.146} \; w_{t} + \underset{(0.753)}{1.796} \; \textrm{ln}w_{p} + \underset{(0.505)}{2.190} \;  y\bigg), \; \; \textrm{if} \; s^{t}=1
	\end{array}
\end{equation*}
From the equations, the son's parents assign him 0.146 extra units from the household total income for an additional unit in his wage. He obtains 1.190 extra units for every unit of nonlabor income in the household. These additional units may appear considerable, but keep in mind that the teen's wage and the household's nonlabor income are modest compared with the parents' wage. In exchange, the parents transfer 0.1796 units (0.1 log point) for every 10\% rise in their wages. Finally, if the son studies, he receives 82.1\% of the total income he receives if he works.

Regarding the home production function, Table \ref{tab:home_production} shows the results. The daughter has a productivity parameter in the public good production of 0.045 units, while the sons have one of 0.038. This difference in productivity might imply an early-age work specialization where daughters are expected to work more in the household than their male counterparts. In contrast, sons work in the labor market instead. However, these results are not as significant as those from the descriptive statistics shown above with daughters providing more domestic hours per week than their male counterparts. The Cobb Douglas assumption on the production function could explain the brief discrepancy in productivity statistics. Because this technology has a substitution parameter equal to one. By relaxing this assumption, I might find a larger difference between daughters and sons. However, to do so, I need to find production shifters that work well with my sample.

\section{Conclusion}
Teens are key members of the household. However, the majority of analyses of them as decision-makers in households have only looked at consumption models. In this study, I show that they play a role in how households allocate time and money. First, I examine the effects of the Costa Rican conditional cash transfer \textit{Avancemos} on the parents' and teens' time allocation. Because the transfer is endogenous to the household's decision-making, I estimate the marginal treatment effect of the teen's education decision, the father's labor supply, the mother's employment status, and the members' domestic work for those households that are unsure whether to apply or not. I find that when parents of daughters receive the transfer, they reduce their working hours and increase their leisure time, indicating a standard income effect. In households with a son, the only effects are on him, who is more likely to attend high school, and his mother, who decreases her labor market involvement. I explain that this gender difference arises from sons bargaining with their parents but not daughters. I provide a collective home model that allows for flexible time allocation between schooling, labor market work, and domestic work to investigate the existence of this bargaining. My findings reject the daughter, but not the son, as a decision-maker, which is consistent with the marginal treatment effect findings.

These findings have significant implications for public policy formulation. Because sons bargain with their parents, they consider the potential wage and the percentage of income allocated to them. Subsidies that are relatively low to encourage sons to stay in school may be ineffective since they do not compensate for the loss in resource sharing that he would receive if he worked. In addition, daughters complement their education with more domestic work, which lowers their opportunity costs to attend school. These two potential mechanisms could also lead to sons specializing in labor and daughters specializing in domestic work. This might explain why, in their adult ages, women have less household bargaining power and lower participation in the labor market.  Public policies aimed at closing gender gaps must consider those gender roles starting at a young age in households.

There is plenty of room for future investigation. It would be interesting to learn who the teen steals bargaining power from. This, however, requires recognizing the mother and father as separate decision-makers. Even if the literature can easily be expanded to cover this concept, the estimation of such models has strong data requirements. It requires enough wage and labor supply for each decision-maker. Additionally, for two of the members an exogenous variation, such as a CCT, without any interaction with each other. The same research could be done on family members, such as grandparents, and other children living in the home. It would also be necessary to examine how much the work of teens, especially daughters, influences home production. Important gender differences resulting from work specialization can be highlighted by this analysis, which may also help to explain a significant portion of women's low labor participation rates.

	\newpage
	\bibliographystyle{apa}
	\bibliography{paper_three.bib}

\newcommand{\noop}[1]{} \DeclareRobustCommand{\firstsecond}[2]{#2}
\begin{thebibliography}{}

\bibitem[\protect\astroncite{Andresen}{2018}]{andresen2018exploring}
Andresen, M.~E. (2018).
\newblock Exploring marginal treatment effects: {F}lexible estimation using
  stata.
\newblock {\em The Stata Journal}, 18(1):118--158.

\bibitem[\protect\astroncite{Ashraf et~al.}{2020}]{ashraf2020negotiating}
Ashraf, N., Bau, N., Low, C., and McGinn, K. (2020).
\newblock Negotiating a better future: How interpersonal skills facilitate
  intergenerational investment.
\newblock {\em The Quarterly Journal of Economics}, 135(2):1095--1151.

\bibitem[\protect\astroncite{Attanasio and
  Lechene}{2014}]{attanasio2014efficient}
Attanasio, O.~P. and Lechene, V. (2014).
\newblock Efficient responses to targeted cash transfers.
\newblock {\em Journal of political Economy}, 122(1):178--222.

\bibitem[\protect\astroncite{Blundell et~al.}{2007}]{blundell2007collective}
Blundell, R., Chiappori, P.-A., Magnac, T., and Meghir, C. (2007).
\newblock Collective labour supply: {H}eterogeneity and non-participation.
\newblock {\em The Review of Economic Studies}, 74(2):417--445.

\bibitem[\protect\astroncite{Blundell et~al.}{2005}]{blundell2005collective}
Blundell, R., Chiappori, P.-A., and Meghir, C. (2005).
\newblock Collective labor supply with children.
\newblock {\em Journal of political Economy}, 113(6):1277--1306.

\bibitem[\protect\astroncite{Blundell and
  Costa~Dias}{2009}]{blundell2009alternative}
Blundell, R. and Costa~Dias, M. (2009).
\newblock Alternative approaches to evaluation in empirical microeconomics.
\newblock {\em Journal of Human Resources}, 44(3):565--640.

\bibitem[\protect\astroncite{Blundell and
  Powell}{2003}]{blundell2003endogeneity}
Blundell, R. and Powell, J.~L. (2003).
\newblock Endogeneity in nonparametric and semiparametric regression models.
\newblock {\em Econometric society monographs}, 36:312--357.

\bibitem[\protect\astroncite{Bonnal et~al.}{1997}]{bonnal1997evaluating}
Bonnal, L., Foug{\`e}re, D., and S{\'e}randon, A. (1997).
\newblock Evaluating the impact of {F}rench employment policies on individual
  labour market histories.
\newblock {\em The Review of Economic Studies}, 64(4):683--713.

\bibitem[\protect\astroncite{Brown et~al.}{2018}]{brown2018sharing}
Brown, C., Calvi, R., and Penglase, J. (2018).
\newblock Sharing the pie: {U}ndernutrition, intra-household allocation, and
  poverty.
\newblock {\em Intra-Household Allocation, and Poverty (June 1, 2018)}.

\bibitem[\protect\astroncite{Calvi}{2020}]{calvi2020older}
Calvi, R. (2020).
\newblock Why are older women missing in {I}ndia? {T}he age profile of
  bargaining power and poverty.
\newblock {\em Journal of Political Economy}, 128(7):2453--2501.

\bibitem[\protect\astroncite{Calvi et~al.}{2021}]{calvi2021more}
Calvi, R., Penglase, J., Tommasi, D., and Wolf, A. (2021).
\newblock The more the poorer? {R}esource sharing and scale economies in large
  families.
\newblock {\em CEPR Discussion Paper No. DP15924}.

\bibitem[\protect\astroncite{Carneiro et~al.}{2011}]{carneiro2011estimating}
Carneiro, P., Heckman, J.~J., and Vytlacil, E.~J. (2011).
\newblock Estimating marginal returns to education.
\newblock {\em American Economic Review}, 101(6):2754--81.

\bibitem[\protect\astroncite{Cherchye et~al.}{2012}]{cherchye2012married}
Cherchye, L., De~Rock, B., and Vermeulen, F. (2012).
\newblock Married with children: {A} collective labor supply model with
  detailed time use and intrahousehold expenditure information.
\newblock {\em American Economic Review}, 102(7):3377--3405.

\bibitem[\protect\astroncite{Chiappori}{1992}]{chiappori1992collective}
Chiappori, P.-A. (1992).
\newblock Collective labor supply and welfare.
\newblock {\em Journal of political Economy}, 100(3):437--467.

\bibitem[\protect\astroncite{Chiappori and
  Mazzocco}{2017}]{chiappori2017static}
Chiappori, P.-A. and Mazzocco, M. (2017).
\newblock Static and intertemporal household decisions.
\newblock {\em Journal of Economic Literature}, 55(3):985--1045.

\bibitem[\protect\astroncite{Cornelissen et~al.}{2016}]{cornelissen2016late}
Cornelissen, T., Dustmann, C., Raute, A., and Sch{\"o}nberg, U. (2016).
\newblock From late to mte: {A}lternative methods for the evaluation of policy
  interventions.
\newblock {\em Labour Economics}, 41:47--60.

\bibitem[\protect\astroncite{Cunha et~al.}{2006}]{cunha2006interpreting}
Cunha, F., Heckman, J.~J., Lochner, L., and Masterov, D.~V. (2006).
\newblock Interpreting the evidence on life cycle skill formation.
\newblock {\em Handbook of the Economics of Education}, 1:697--812.

\bibitem[\protect\astroncite{Dauphin et~al.}{2011}]{dauphin2011children}
Dauphin, A., El~Lahga, A.-R., Fortin, B., and Lacroix, G. (2011).
\newblock Are children decision-makers within the household?
\newblock {\em The Economic Journal}, 121(553):871--903.

\bibitem[\protect\astroncite{De~Hoyos et~al.}{2016}]{de2016out}
De~Hoyos, R., Rogers, H., and Sz{\'e}kely, M. (2016).
\newblock Out of school and out of work: {R}isk and opportunities for latin
  america’s ninis.

\bibitem[\protect\astroncite{De~Rock et~al.}{2022}]{de2022household}
De~Rock, B., Potoms, T., and Tommasi, D. (2022).
\newblock Household responses to cash transfers.
\newblock {\em Economic development and cultural change}, 70(2):625--652.

\bibitem[\protect\astroncite{Del~Boca et~al.}{2014}]{del2014household}
Del~Boca, D., Flinn, C., and Wiswall, M. (2014).
\newblock Household choices and child development.
\newblock {\em Review of Economic Studies}, 81(1):137--185.

\bibitem[\protect\astroncite{DiPasquale and
  Glaeser}{1999}]{dipasquale1999incentives}
DiPasquale, D. and Glaeser, E.~L. (1999).
\newblock Incentives and social capital: {A}re homeowners better citizens?
\newblock {\em Journal of urban Economics}, 45(2):354--384.

\bibitem[\protect\astroncite{Dubois and Rubio-Codina}{2012}]{dubois2012child}
Dubois, P. and Rubio-Codina, M. (2012).
\newblock Child care provision: {S}emiparametric evidence from a randomized
  experiment in mexico.
\newblock {\em Annals of Economics and Statistics/ANNALES D'{\'E}CONOMIE ET DE
  STATISTIQUE}, pages 155--184.

\bibitem[\protect\astroncite{Dunbar et~al.}{2013}]{dunbar2013children}
Dunbar, G.~R., Lewbel, A., and Pendakur, K. (2013).
\newblock Children's resources in collective households: {I}dentification,
  estimation, and an application to child poverty in {M}alawi.
\newblock {\em American Economic Review}, 103(1):438--71.

\bibitem[\protect\astroncite{Garc{\'\i}a and
  Saavedra}{2017}]{garcia2017educational}
Garc{\'\i}a, S. and Saavedra, J.~E. (2017).
\newblock Educational impacts and cost-effectiveness of conditional cash
  transfer programs in developing countries: A meta-analysis.
\newblock {\em Review of Educational Research}, 87(5):921--965.

\bibitem[\protect\astroncite{Heckman and
  Vytlacil}{2005}]{heckman2005structural}
Heckman, J.~J. and Vytlacil, E. (2005).
\newblock Structural equations, treatment effects, and econometric policy
  evaluation 1.
\newblock {\em Econometrica}, 73(3):669--738.

\bibitem[\protect\astroncite{Hern{\'a}ndez~Romero}{2016}]{hernandez2016como}
Hern{\'a}ndez~Romero, K. (2016).
\newblock ?`{C}{\'o}mo funciona {A}vancemos?: {M}ejores pr{\'a}cticas en la
  implementaci{\'o}n de programas de transferencias monetarias condicionadas en
  {A}m{\'e}rica {L}atina y el {C}aribe.
\newblock {\em BID}.

\bibitem[\protect\astroncite{{{I}nstituto {M}ixto de {A}yuda {S}ocial de
  {C}osta {R}ica}}{2022}]{imasweb}
{{I}nstituto {M}ixto de {A}yuda {S}ocial de {C}osta {R}ica} (2022).
\newblock Avancemos.
\newblock Accessed: 22-02-2022.

\bibitem[\protect\astroncite{{{I}nstituto {N}cional de {E}stad{\'i}sticas y
  {C}ensos}}{2015}]{enaho2015}
{{I}nstituto {N}cional de {E}stad{\'i}sticas y {C}ensos} (2015).
\newblock Canasta b{\'a}sica alimentaria - julio 2015.
\newblock Accessed: 26-04-2022.

\bibitem[\protect\astroncite{{{I}nstituto {N}cional de {E}stad{\'i}sticas y
  {C}ensos}}{2022}]{enaho}
{{I}nstituto {N}cional de {E}stad{\'i}sticas y {C}ensos} (2022).
\newblock Encuesta {N}acional de {H}ogares.
\newblock Accessed: 30-03-2020.

\bibitem[\protect\astroncite{Jim{\'e}nez-Fontana}{2015}]{jimenez2015analysis}
Jim{\'e}nez-Fontana, P. (2015).
\newblock Analysis of non-remunerated production in {C}osta {R}ica.
\newblock {\em The Journal of the Economics of Ageing}, 5:45--53.

\bibitem[\protect\astroncite{Keshavarz~Haddad}{2017}]{keshavarz2017parents}
Keshavarz~Haddad, G. (2017).
\newblock Parents’ decision on child labour and school attendance: {E}vidence
  from {I}ranian households.
\newblock {\em Journal of Education and Work}, 30(6):612--631.

\bibitem[\protect\astroncite{Lang~Clachar et~al.}{2015}]{lang2015efecto}
Lang~Clachar, G., Soto~M{\'e}ndez, B., and Robalino, J. (2015).
\newblock Efecto del programa ``{A}vancemos” sobre el trabajo infantil en
  {C}osta {R}ica.
\newblock Technical report.

\bibitem[\protect\astroncite{Lundberg et~al.}{2009}]{lundberg2009decision}
Lundberg, S., Romich, J.~L., and Tsang, K.~P. (2009).
\newblock Decision-making by children.
\newblock {\em Review of Economics of the Household}, 7(1):1--30.

\bibitem[\protect\astroncite{Mata and Hern{\'a}ndez}{2015}]{mata2015evaluacion}
Mata, C. and Hern{\'a}ndez, K. (2015).
\newblock Evaluaci{\'o}n de impacto de la implementaci{\'o}n de transferencias
  monetarias condicionadas para educaci{\'o}n secundaria en {C}osta {R}ica
  ({A}vancemos).
\newblock {\em Revista de Ciencias Econ{\'o}micas}, 33(1):9--35.

\bibitem[\protect\astroncite{Meza-Cordero}{2011}]{meza2011effects}
Meza-Cordero, J.~A. (2011).
\newblock The effects of subsidizing secondary schooling: {E}vidence from a
  conditional cash transfer program in {C}osta {R}ica.
\newblock {\em Unpublished Document, University of Southern California}.

\bibitem[\protect\astroncite{Mu{\~n}oz-Alvarado}{2016}]{munoz2016avancemos}
Mu{\~n}oz-Alvarado, J.~A. (2016).
\newblock Avancemos: {E}fectos sobre el abandono educativo en el hogar de
  personas beneficiarias.
\newblock {\em Revista Electr{\'o}nica Educare}, 20(1):53--74.

\bibitem[\protect\astroncite{Reggio}{2011}]{reggio2011influence}
Reggio, I. (2011).
\newblock The influence of the mother's power on her child's labor in {M}exico.
\newblock {\em Journal of development economics}, 96(1):95--105.

\bibitem[\protect\astroncite{Sokullu and Valente}{2022}]{sokullu2022individual}
Sokullu, S. and Valente, C. (2022).
\newblock Individual consumption in collective households: {I}dentification
  using repeated observations with an application to {P}rogresa.
\newblock {\em Journal of Applied Econometrics}, 37(2):286--304.

\bibitem[\protect\astroncite{Stern}{1986}]{stern1986specification}
Stern, N. (1986).
\newblock On the specification of labour supply functions.
\newblock In Blundell, R.~W., Blundell, R., and Walker, I., editors, {\em
  Unemployment, search and labour supply}. Cambridge University Press.

\bibitem[\protect\astroncite{Todd and Wolpin}{2006}]{todd2006assessing}
Todd, P.~E. and Wolpin, K.~I. (2006).
\newblock Assessing the impact of a school subsidy program in mexico: {U}sing a
  social experiment to validate a dynamic behavioral model of child schooling
  and fertility.
\newblock {\em American economic review}, 96(5):1384--1417.

\bibitem[\protect\astroncite{Tommasi}{2019}]{tommasi2019control}
Tommasi, D. (2019).
\newblock Control of resources, bargaining power and the demand of food:
  {E}vidence from {P}rogresa.
\newblock {\em Journal of Economic Behavior \& Organization}, 161:265--286.

\bibitem[\protect\astroncite{Vermeulen}{2002}]{vermeulen2002collective}
Vermeulen, F. (2002).
\newblock Collective household models: {P}rinciples and main results.
\newblock {\em Journal of Economic Surveys}, 16(4):533--564.

\end{thebibliography}
	
	\newpage
	\appendix

	\setcounter{table}{0}
	\renewcommand{\thetable}{A\arabic{table}}
	\renewcommand*{\theHtable}{\thetable}
	
	\setcounter{figure}{0}
	\renewcommand{\thefigure}{A\arabic{figure}}
	\renewcommand*{\theHfigure}{\thefigure}
	
	\setcounter{equation}{0}
	\renewcommand{\theequation}{A\arabic{equation}}
	\renewcommand*{\theHequation}{\theequation}
		
	\section{Reduced form results}\label{app:mte_results}
		
\begin{figure}[h!]
			\centering
		\caption{Propensity scores for receiving \textit{Avancemos}}%
		\label{fig:pscores}%
			\begin{subfigure}{.45\linewidth}
				\centering\includegraphics[width=1\columnwidth]{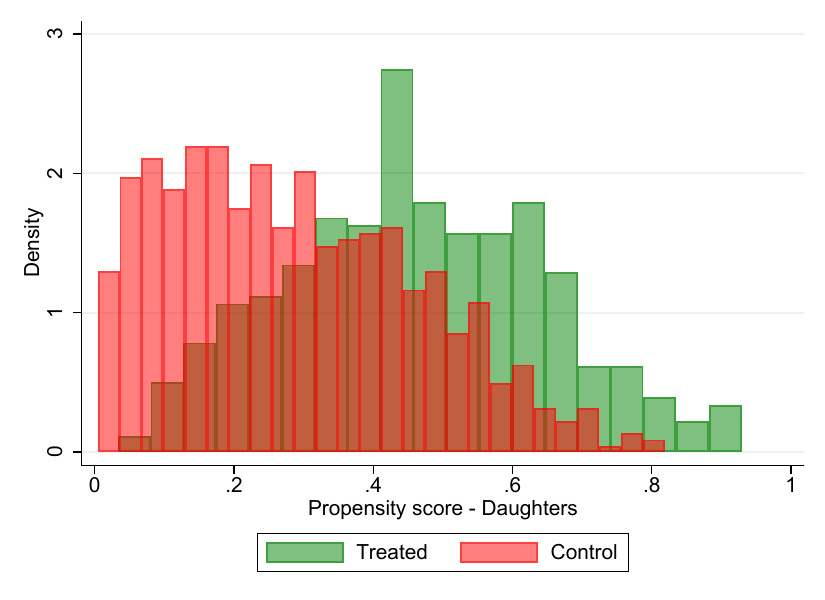}
				\caption{Daughters.}
			\end{subfigure}\hspace{10mm}
			\begin{subfigure}{.45\linewidth}
				\centering\includegraphics[width=1\columnwidth]{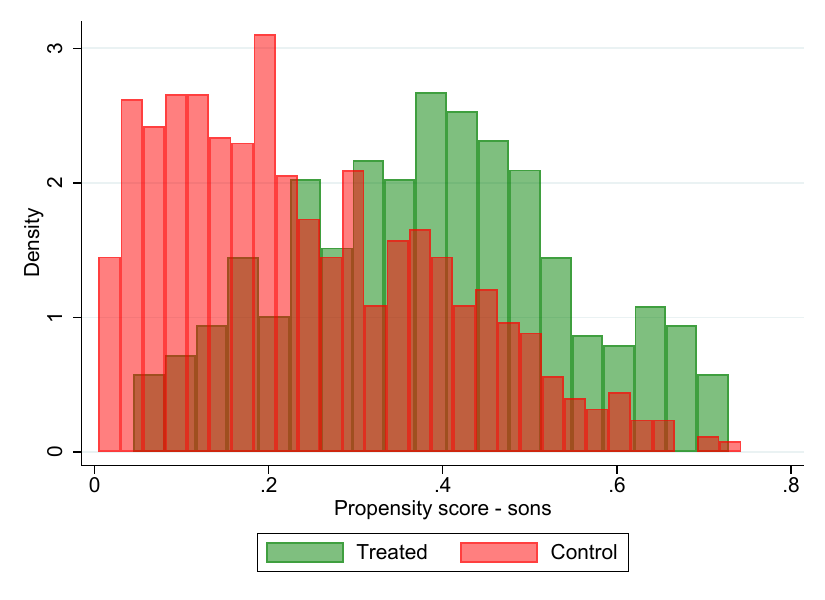}
				\caption{Sons.}
			\end{subfigure}
		\end{figure}

		Table \ref{tab:pscores} shows the results of the first stage of the MTE estimation for the three samples: complete, daughters and sons. Figure \ref{fig:pscores} shows the graphs of the propensity scores. I estimate them using a probit. 
		
		\begin{table}[H]\centering 
			\def\sym#1{\ifmmode^{#1}\else\(^{#1}\)\fi}
			\caption{First stage: \textit{Avancemos}' marginal effects}
			\label{tab:pscores} 
			\begin{adjustbox}{width=11cm,totalheight=23cm}
				\begin{threeparttable}
\begin{tabular}{l*{2}{c}}
	\hline\hline
	&\multicolumn{1}{c}{Daughters}&\multicolumn{1}{c}{Sons}\\
	&\multicolumn{1}{c}{} &\multicolumn{1}{c}{} \\
	\hline
Percentage of households with \textit{Avancemos}   &       2.138\sym{***}&       1.449\sym{***}\\
	&     (0.271)         &     (0.256)         \\
	[1em]
	Father years of schooling&    0.000128         &     0.00448         \\
	&   (0.00678)         &   (0.00580)         \\
	[1em]
	Father high school or more diploma&     -0.0646         &      -0.133\sym{**} \\
	&    (0.0575)         &    (0.0500)         \\
	[1em]
	Father occupation mechanic&       0.122\sym{***}&      0.0138         \\
	&    (0.0328)         &    (0.0288)         \\
	[1em]
	Father occupation elemental&       0.135\sym{***}&      0.0336         \\
	&    (0.0338)         &    (0.0292)         \\
	[1em]
	Father occupation services and transport&     -0.0588\sym{*}  &     -0.0392         \\
	&    (0.0297)         &    (0.0253)         \\
	[1em]
	Father insurance employed&     -0.0506         &    -0.00675         \\
	&    (0.0275)         &    (0.0237)         \\
	[1em]
	Mother years of schooling&    -0.00363         &    -0.00646         \\
	&   (0.00654)         &   (0.00622)         \\
	[1em]
	Mother high school or more diploma&      -0.189\sym{**} &     -0.0890         \\
	&    (0.0586)         &    (0.0514)         \\
	[1em]
	Teen age 18 or more &      -0.239\sym{***}&      -0.221\sym{***}\\
	&    (0.0363)         &    (0.0275)         \\
	[1em]
	Age father          &    0.000264         &    -0.00101         \\
	&   (0.00229)         &   (0.00215)         \\
	[1em]
	Age mother          &   -0.000192         &   -0.000141         \\
	&   (0.00293)         &   (0.00260)         \\
	[1em]
	One younger children&       0.120\sym{***}&       0.178\sym{***}\\
	&    (0.0363)         &    (0.0296)         \\
	[1em] 
	\hline
	Year effects&     Yes        &     Yes         \\
	Geographical effects&    Yes        &    Yes         \\
	Observations        &        1,090         &        1,357         \\
	\(R^{2}\)           &         0.146            &      0.127               \\
						\hline\hline
					\end{tabular}
					\begin{tablenotes}\footnotesize
						\item[] \footnotesize Standard errors between parenthesis. \sym{*} \(p<0.05\), \sym{**} \(p<0.01\), \sym{***} \(p<0.001\).
						\item[] \footnotesize Baseline categories: no education diploma or school diploma for parents, different occupations (manager, research, technical and academic professors, staff and agriculture), different industries (mines and agriculture, finance, public administration, real state, teaching, social health, domestic and others), no insurance or non-employed insurance for the father, no younger children in the household.
					\end{tablenotes}
				\end{threeparttable}
			\end{adjustbox}
		\end{table}

\begin{figure}[h!]
			\centering
			\caption{\textit{Avancemos} on household with teen daughter: MTE.}%
			\label{fig:daughters}%
			\begin{subfigure}{.45\linewidth}
				\centering\includegraphics[width=1\columnwidth]{mte_school_daughters.pdf}
				\caption{Teen's schooling decision.}
			\end{subfigure}\hspace{10mm}
			\begin{subfigure}{.45\linewidth}
				\centering\includegraphics[width=1\columnwidth]{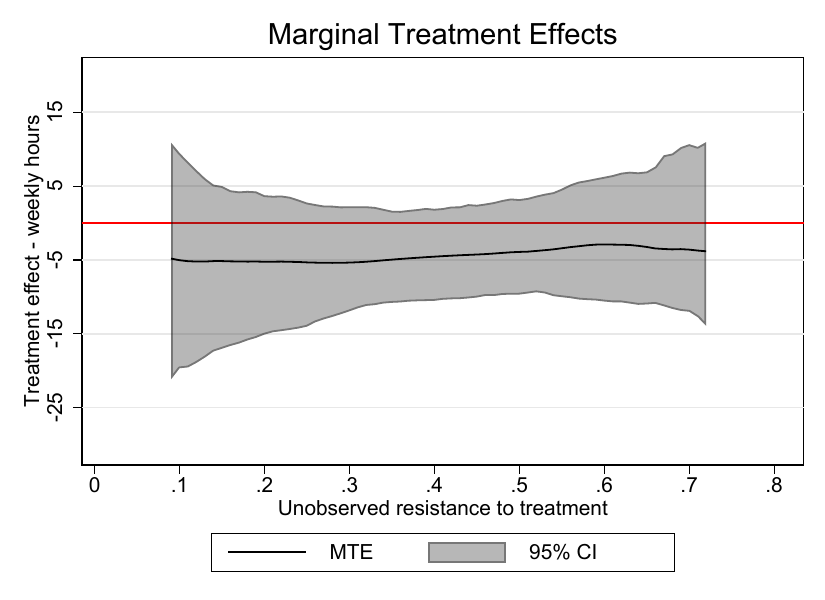}
				\caption{Teen's domestic labor supply.}
			\end{subfigure}
			\begin{subfigure}{.45\linewidth}
				\centering\includegraphics[width=1\columnwidth]{mte_labor_hours_father_daughters.pdf}
			\caption{Father's market labor supply.}
		\end{subfigure}\hspace{10mm}
		\begin{subfigure}{.45\linewidth}
			\centering\includegraphics[width=1\columnwidth]{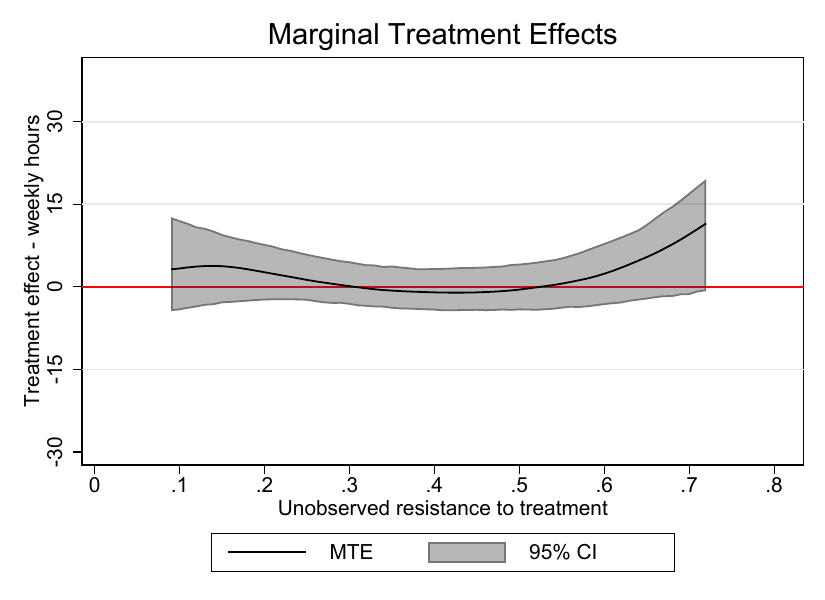}
			\caption{Father's domestic supply.}
		\end{subfigure}
		\begin{subfigure}{.45\linewidth}
			\centering\includegraphics[width=1\columnwidth]{mte_employed_mother_daughters.pdf}
			\caption{Mother's employment status.}
		\end{subfigure}\hspace{10mm}
		\begin{subfigure}{.45\linewidth}
			\centering\includegraphics[width=1\columnwidth]{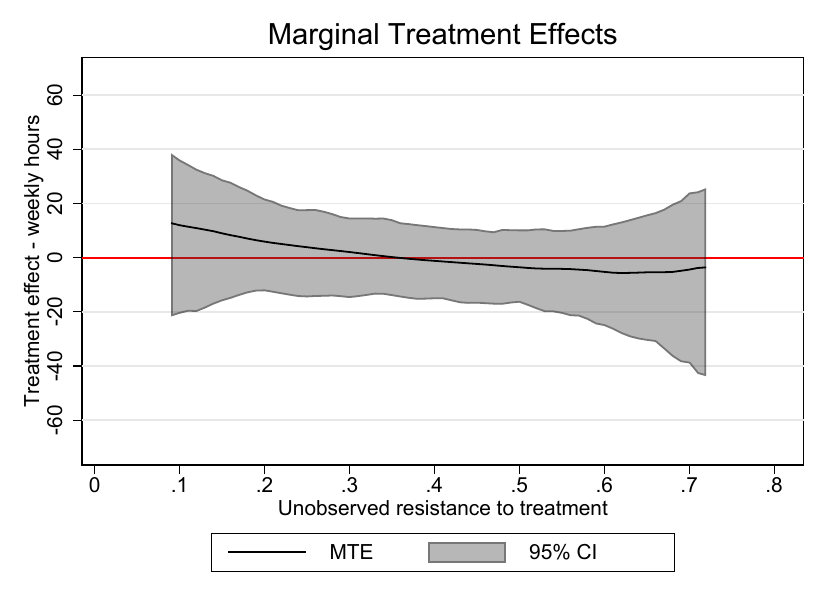}
			\caption{Mother's domestic supply.}
		\end{subfigure}
		\end{figure}

		Table \ref{tab:daughters_mte} shows the estimation results of observables for the MTE for the daughters' sample and Figure \ref{fig:daughters} shows the graphs of the estimated unobservable.
		
		\begin{landscape}
			\begin{table}[htbp]\centering
				\def\sym#1{\ifmmode^{#1}\else\(^{#1}\)\fi}
				\caption{\textit{Avancemos'} effect on households for the daughters sample: MTE.}
				\label{tab:daughters_mte}
				\begin{adjustbox}{width=26cm,totalheight=17cm} 
					\begin{threeparttable}
						\begin{tabular}{l*{12}{c}}
							\hline\hline
							&\multicolumn{2}{c}{Teen's schooling}      &\multicolumn{2}{c}{Teen's domestic hours}     &\multicolumn{2}{c}{Mother's employment}        &\multicolumn{2}{c}{Mother's domestic hours}      &\multicolumn{2}{c}{Father's market hours}      &\multicolumn{2}{c}{Father's domestic hours}     \\
							&        Baseline         & Difference if treated         &        Baseline         & Difference if treated         &        Baseline         & Difference if treated         &        Baseline         & Difference if treated         &        Baseline         & Difference if treated         &        Baseline         & Difference if treated         \\
							\hline
							Father years of schooling&       0.015\sym{*}  &      -0.015\sym{*}  &      -0.421\sym{*}  &       0.327         &       0.001         &       0.003         &       0.009         &       1.518\sym{*}  &       0.095         &       0.320         &       0.173         &       0.109         \\
							&     (0.007)         &     (0.007)         &     (0.192)         &     (0.322)         &     (0.009)         &     (0.016)         &     (0.388)         &     (0.671)         &     (0.229)         &     (0.373)         &     (0.104)         &     (0.189)         \\
							[1em]
							Father high school or more diploma&      -0.050         &       0.050         &      -0.190         &      -1.706         &      -0.027         &      -0.052         &       0.006         &     -11.123\sym{*}  &      -3.338\sym{*}  &      -3.037         &      -0.134         &      -0.881         \\
							&     (0.055)         &     (0.055)         &     (1.459)         &     (2.236)         &     (0.071)         &     (0.124)         &     (2.962)         &     (5.496)         &     (1.691)         &     (3.132)         &     (0.887)         &     (1.476)         \\
							[1em]
							Father occupation elemental&      -0.005         &       0.005         &       1.204         &      -2.119         &       0.161\sym{**} &      -0.182\sym{*}  &      -4.863\sym{*}  &       3.699         &      -1.760         &       0.272         &       0.368         &      -0.554         \\
							&     (0.051)         &     (0.051)         &     (1.063)         &     (1.603)         &     (0.052)         &     (0.077)         &     (2.102)         &     (3.503)         &     (1.404)         &     (2.229)         &     (0.729)         &     (1.018)         \\
							[1em]
							Father occupation services and transport&      -0.026         &       0.026         &      -0.155         &       1.206         &       0.005         &       0.078         &      -3.496\sym{*}  &       1.839         &       1.449         &       2.129         &      -0.475         &       0.973         \\
							&     (0.030)         &     (0.030)         &     (0.889)         &     (1.476)         &     (0.041)         &     (0.072)         &     (1.636)         &     (2.875)         &     (0.977)         &     (1.873)         &     (0.523)         &     (0.814)         \\
							[1em]
							Father insurance employed&       0.002         &      -0.002         &      -0.489         &       1.132         &      -0.095\sym{*}  &       0.074         &       2.151         &      -4.057         &       3.010\sym{**} &       3.172         &       0.029         &       0.319         \\
							&     (0.035)         &     (0.035)         &     (0.893)         &     (1.313)         &     (0.039)         &     (0.062)         &     (1.580)         &     (2.763)         &     (0.982)         &     (1.639)         &     (0.457)         &     (0.728)         \\
							[1em]
							Mother years of schooling&       0.025\sym{***}&      -0.025\sym{***}&      -0.160         &       0.399         &       0.021\sym{*}  &      -0.012         &      -0.105         &      -0.386         &      -0.099         &      -0.480         &       0.332\sym{**} &      -0.385\sym{*}  \\
							&     (0.007)         &     (0.007)         &     (0.168)         &     (0.304)         &     (0.009)         &     (0.016)         &     (0.354)         &     (0.674)         &     (0.231)         &     (0.372)         &     (0.124)         &     (0.194)         \\
							[1em]
							Mother high school or more diploma&      -0.097         &       0.097         &       0.063         &       0.437         &       0.112         &       0.002         &      -4.599         &       5.650         &      -2.231         &       3.936         &       0.980         &      -0.293         \\
							&     (0.056)         &     (0.056)         &     (1.635)         &     (2.733)         &     (0.077)         &     (0.149)         &     (3.156)         &     (6.276)         &     (1.821)         &     (3.405)         &     (0.946)         &     (1.367)         \\
							[1em]
							Teen age 18 or more &      -0.322\sym{***}&       0.322\sym{***}&       4.326\sym{***}&      -0.678         &      -0.118         &       0.065         &      -2.110         &       3.542         &      -3.222\sym{*}  &       0.414         &       0.752         &      -1.443         \\
							&     (0.054)         &     (0.054)         &     (1.290)         &     (2.342)         &     (0.060)         &     (0.104)         &     (2.408)         &     (4.815)         &     (1.386)         &     (2.973)         &     (0.681)         &     (1.174)         \\
							[1em]
							Age mother          &      -0.000         &       0.000         &      -0.003         &       0.046         &      -0.004         &      -0.004         &       0.237         &       0.001         &      -0.255\sym{*}  &       0.096         &       0.041         &      -0.048         \\
							&     (0.003)         &     (0.003)         &     (0.084)         &     (0.139)         &     (0.004)         &     (0.008)         &     (0.156)         &     (0.300)         &     (0.105)         &     (0.193)         &     (0.048)         &     (0.084)         \\
							[1em]
							One younger children&       0.024         &      -0.024         &       0.425         &      -2.316         &      -0.072         &       0.039         &       9.052\sym{***}&      -2.625         &       0.065         &      -1.432         &      -0.033         &      -0.886         \\
							&     (0.038)         &     (0.038)         &     (1.036)         &     (1.808)         &     (0.049)         &     (0.087)         &     (1.987)         &     (3.548)         &     (1.172)         &     (2.291)         &     (0.646)         &     (1.018)         \\
							[1em]
							Two or more younger children&      -0.020         &       0.020         &       1.712         &      -3.905\sym{*}  &      -0.125\sym{*}  &       0.034         &      16.746\sym{***}&      -6.177         &       0.299         &       0.510         &       0.733         &      -1.457         \\
							&     (0.050)         &     (0.050)         &     (1.128)         &     (1.954)         &     (0.054)         &     (0.096)         &     (2.369)         &     (4.105)         &     (1.458)         &     (2.558)         &     (0.751)         &     (1.144)         \\
							[1em]
							Constant            &       0.475\sym{**} &       0.525\sym{**} &      15.941\sym{***}&      -0.439         &       0.432         &       0.088         &      22.306\sym{**} &       0.760         &      65.499\sym{***}&     -10.036         &      -3.764         &       4.958         \\
							&     (0.175)         &     (0.175)         &     (4.400)         &     (7.401)         &     (0.227)         &     (0.370)         &     (8.634)         &    (14.303)         &     (6.257)         &     (9.391)         &     (2.930)         &     (4.460)         \\
						[1em]
						\hline
						Year and geographical effects &\multicolumn{2}{c}{Yes}      &\multicolumn{2}{c}{Yes}     &\multicolumn{2}{c}{Yes}        &\multicolumn{2}{c}{Yes}      &\multicolumn{2}{c}{Yes}      &\multicolumn{2}{c}{Yes}     \\
						[1em]
						P-value observable heterogeneity &\multicolumn{2}{c}{ 0.0000 }      &\multicolumn{2}{c}{ 0.6466 }     &\multicolumn{2}{c}{ 0.0293 }        &\multicolumn{2}{c}{ 0.1857  }      &\multicolumn{2}{c}{ 0.4465 }      &\multicolumn{2}{c}{ 0.0481}     \\
						[1em]
						P-value no unobservable heterogeneity &\multicolumn{2}{c}{  0.991 }      &\multicolumn{2}{c}{ 0.9993 }     &\multicolumn{2}{c}{ 0.8920 }        &\multicolumn{2}{c}{ 0.9996 }      &\multicolumn{2}{c}{0.8150 }      &\multicolumn{2}{c}{ 0.8719 }     \\
						[1em]
						Observations        &        \multicolumn{2}{c}{ 1,090}                           &        \multicolumn{2}{c}{1,090}                   &        \multicolumn{2}{c}{ 1,090}                          &        \multicolumn{2}{c}{ 1,090}                           &        \multicolumn{2}{c}{ 1,090}                            &        \multicolumn{2}{c}{ 1,090}                    \\
						\hline\hline
					\end{tabular}
						\begin{tablenotes}\footnotesize
							\item[] \footnotesize \sym{*} \(p<0.05\), \sym{**} \(p<0.01\), \sym{***} \(p<0.001\).
							\item[] \footnotesize Baseline categories: no education diploma for parents, different occupations (manager, research, technical and academic professors and staff), different industries (finance, public administration, real state, teaching, social health, domestic and others), no insurance, no younger children in the household and for the geographical variable it is living outside the Central Valley in a rural zone.
							\item[] Omitted variables from the table because of their non significant effect: father's age, mother's high school diploma or more, indicator variable for father with a mechanic occupation.
							\item[] The reported test statistic for observable heterogeneity tests for the joint significance of all elements in the column "Difference if treated", while the test for unobservable heterogeneity tests if MTEs differ with unobserved costs of treatment. I estimated the model in STATA with the user written command ``mtefe" \citep{andresen2018exploring}.
						\end{tablenotes}
					\end{threeparttable}
				\end{adjustbox}
			\end{table}
		\end{landscape}

\begin{figure}[h!]
			\centering
			\caption{\textit{Avancemos} on household with teen son: MTE.}%
			\label{fig:sons}%
			\begin{subfigure}{.45\linewidth}
				\centering\includegraphics[width=1\columnwidth]{mte_school_sons.pdf}
				\caption{Teen's schooling decision.}
			\end{subfigure}\hspace{10mm}
			\begin{subfigure}{.45\linewidth}
				\centering\includegraphics[width=1\columnwidth]{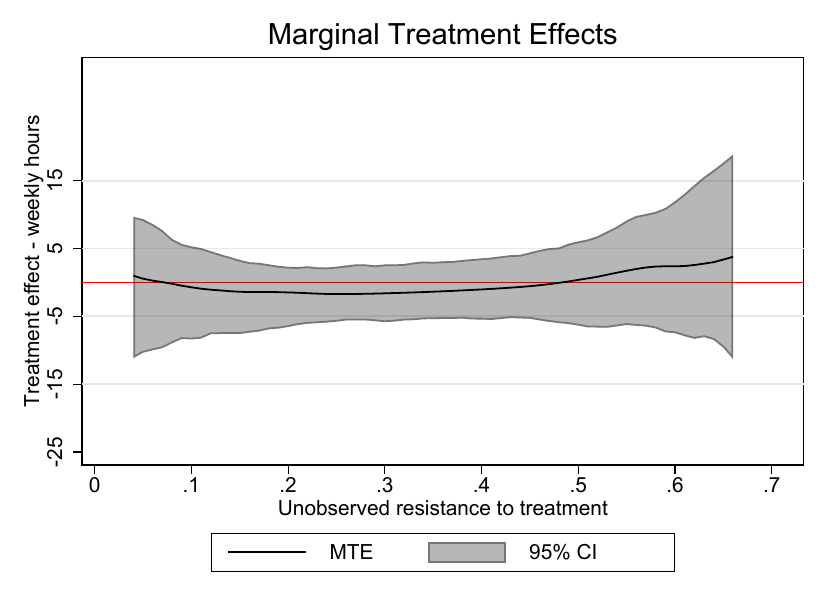}
				\caption{Teen's domestic labor supply.}
			\end{subfigure}
			\begin{subfigure}{.45\linewidth}
				\centering\includegraphics[width=1\columnwidth]{mte_labor_hours_father_sons.pdf}
				\caption{Father's market labor supply.}
			\end{subfigure}\hspace{10mm}
			\begin{subfigure}{.45\linewidth}
				\centering\includegraphics[width=1\columnwidth]{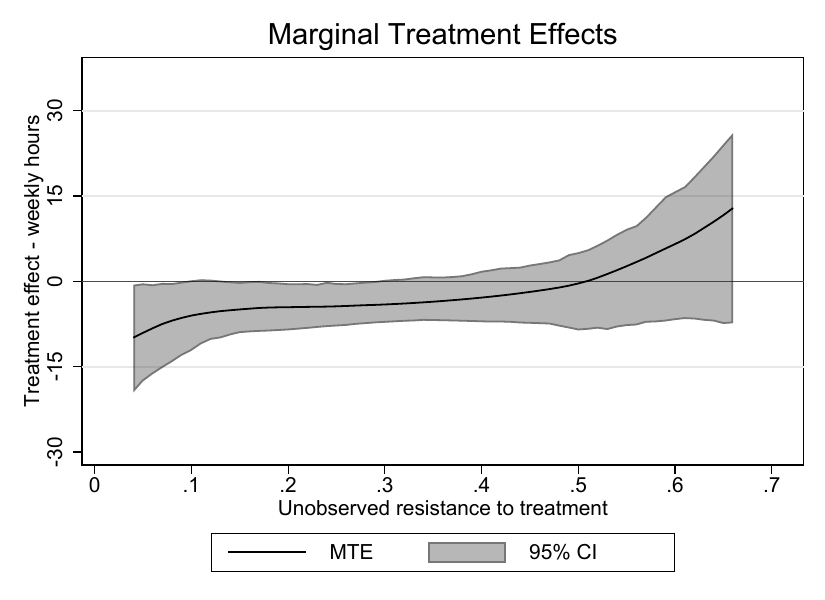}
				\caption{Father's domestic supply.}
			\end{subfigure}
			\begin{subfigure}{.45\linewidth}
				\centering\includegraphics[width=1\columnwidth]{mte_employed_mother_sons.pdf}
				\caption{Mother's employment status.}
			\end{subfigure}\hspace{10mm}
			\begin{subfigure}{.45\linewidth}
				\centering\includegraphics[width=1\columnwidth]{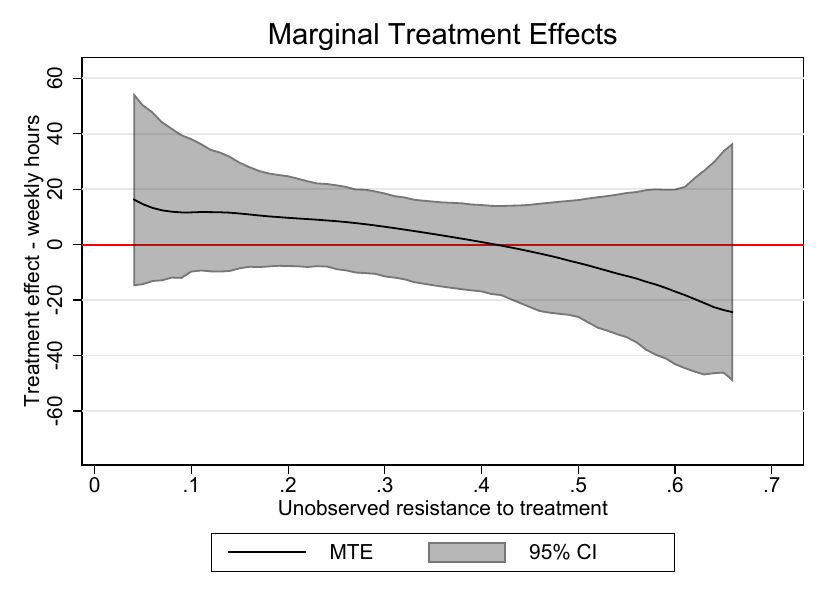}
				\caption{Mother's domestic supply.}
			\end{subfigure}
		\end{figure}

		Table \ref{tab:sons_mte} shows the estimation results of observables for the MTE for the sons' sample and Figure \ref{fig:sons} shows the graphs of the estimated unobservable.

		\begin{landscape}
			\begin{table}[htbp]\centering
				\def\sym#1{\ifmmode^{#1}\else\(^{#1}\)\fi}
				\caption{\textit{Avancemos'} effect on households for the sons sample: MTE.}
				\label{tab:sons_mte}%
				\begin{adjustbox}{width=26cm,totalheight=17cm}
					\begin{threeparttable}
						\begin{tabular}{l*{12}{c}}
							\hline\hline
							&\multicolumn{2}{c}{Teen's schooling}      &\multicolumn{2}{c}{Teen's domestic hours}     &\multicolumn{2}{c}{Mother's employment}        &\multicolumn{2}{c}{Mother's domestic hours}      &\multicolumn{2}{c}{Father's market hours}      &\multicolumn{2}{c}{Father's domestic hours}     \\
							&        Baseline         & Difference if treated         &        Baseline         & Difference if treated         &        Baseline         & Difference if treated         &        Baseline         & Difference if treated         &        Baseline         & Difference if treated         &        Baseline         & Difference if treated         \\
							\hline   
							Father high school or more diploma&       0.085         &      -0.085         &      -0.874         &      -1.523         &       0.019         &      -0.039         &      -0.165         &      -0.720         &      -3.323\sym{*}  &       1.591         &      -0.480         &       0.567         \\
							&     (0.055)         &     (0.055)         &     (0.760)         &     (1.325)         &     (0.070)         &     (0.155)         &     (2.712)         &     (6.043)         &     (1.642)         &     (3.302)         &     (0.795)         &     (1.672)         \\
							[1em]
							Father occupation elemental&      -0.124\sym{**} &       0.124\sym{**} &      -0.228         &       0.832         &      -0.053         &       0.080         &      -0.028         &      -2.066         &      -2.973\sym{**} &       1.371         &       0.046         &       0.010         \\
							&     (0.039)         &     (0.039)         &     (0.436)         &     (0.754)         &     (0.039)         &     (0.066)         &     (1.571)         &     (2.869)         &     (1.015)         &     (1.865)         &     (0.464)         &     (0.849)         \\
							[1em]
							Father occupation services and transport&      -0.065\sym{*}  &       0.065\sym{*}  &      -0.162         &       1.050         &       0.051         &      -0.051         &      -2.255         &       0.206         &       3.125\sym{***}&      -0.004         &      -0.522         &       0.058         \\
							&     (0.029)         &     (0.029)         &     (0.365)         &     (0.694)         &     (0.035)         &     (0.068)         &     (1.307)         &     (2.735)         &     (0.858)         &     (1.810)         &     (0.376)         &     (0.725)         \\
							[1em]
							Father insurance employed&       0.010         &      -0.010         &       0.650         &      -1.014         &       0.046         &      -0.140\sym{*}  &       0.277         &       1.078         &       2.520\sym{***}&       4.068\sym{**} &       0.364         &      -0.358         \\
							&     (0.029)         &     (0.029)         &     (0.383)         &     (0.698)         &     (0.029)         &     (0.059)         &     (1.367)         &     (2.409)         &     (0.686)         &     (1.393)         &     (0.348)         &     (0.723)         \\
							[1em]
							Mother years of schooling&       0.020\sym{**} &      -0.020\sym{**} &       0.051         &       0.089         &       0.025\sym{**} &      -0.016         &      -1.109\sym{**} &       1.154         &      -0.211         &      -0.044         &       0.289\sym{**} &       0.030         \\
							&     (0.008)         &     (0.008)         &     (0.096)         &     (0.184)         &     (0.008)         &     (0.017)         &     (0.356)         &     (0.694)         &     (0.181)         &     (0.414)         &     (0.101)         &     (0.180)         \\
							[1em]
							Mother high school or more diploma&       0.020         &      -0.020         &      -0.285         &       0.402         &      -0.066         &      -0.094         &       2.841         &       0.628         &      -0.519         &       2.873         &      -1.729\sym{*}  &       1.002         \\
							&     (0.056)         &     (0.056)         &     (0.790)         &     (1.754)         &     (0.067)         &     (0.134)         &     (2.768)         &     (5.752)         &     (1.579)         &     (3.479)         &     (0.868)         &     (1.729)         \\
							[1em]
							Teen age 18 or more &      -0.414\sym{***}&       0.414\sym{***}&      -0.842         &       1.868         &      -0.077         &      -0.078         &       0.017         &       0.038         &      -1.325         &       5.110         &      -1.248         &       0.839         \\
							&     (0.047)         &     (0.047)         &     (0.654)         &     (1.275)         &     (0.054)         &     (0.101)         &     (1.964)         &     (4.999)         &     (1.199)         &     (3.078)         &     (0.689)         &     (1.389)         \\
							[1em]
							Age father          &       0.003         &      -0.003         &      -0.032         &      -0.071         &      -0.004         &       0.003         &       0.138         &      -0.606\sym{**} &      -0.086         &      -0.292         &      -0.074\sym{*}  &       0.030         \\
							&     (0.003)         &     (0.003)         &     (0.036)         &     (0.061)         &     (0.003)         &     (0.006)         &     (0.117)         &     (0.221)         &     (0.067)         &     (0.154)         &     (0.034)         &     (0.064)         \\
							[1em]
							Age mother          &      -0.005         &       0.005         &       0.036         &      -0.062         &      -0.007         &       0.004         &       0.333\sym{*}  &       0.048         &      -0.104         &       0.130         &      -0.004         &      -0.010         \\
							&     (0.003)         &     (0.003)         &     (0.043)         &     (0.076)         &     (0.004)         &     (0.007)         &     (0.132)         &     (0.272)         &     (0.086)         &     (0.192)         &     (0.038)         &     (0.079)         \\
							[1em]
							One younger children&       0.044         &      -0.044         &       1.381\sym{*}  &      -1.246         &       0.057         &      -0.098         &       4.531\sym{*}  &       4.229         &       0.031         &      -2.105         &       0.527         &       0.419         \\
							&     (0.043)         &     (0.043)         &     (0.604)         &     (1.218)         &     (0.057)         &     (0.105)         &     (2.151)         &     (4.749)         &     (1.268)         &     (2.708)         &     (0.590)         &     (1.297)         \\
							[1em]
							Two or more younger children&      -0.052         &       0.052         &       1.166\sym{*}  &      -1.334         &      -0.099         &      -0.015         &      12.808\sym{***}&       2.314         &       0.357         &      -2.991         &       1.125         &      -0.258         \\
							&     (0.042)         &     (0.042)         &     (0.515)         &     (1.050)         &     (0.053)         &     (0.102)         &     (2.139)         &     (4.595)         &     (1.182)         &     (2.748)         &     (0.619)         &     (1.219)         \\
							[1em]
							Constant            &       0.697\sym{***}&       0.303         &       2.879         &       4.693         &       0.701\sym{***}&      -0.725         &      18.698\sym{*}  &      24.711         &      59.618\sym{***}&       2.548         &       4.338\sym{*}  &      -2.609         \\
							&     (0.165)         &     (0.165)         &     (1.992)         &     (4.365)         &     (0.183)         &     (0.448)         &     (7.419)         &    (17.844)         &     (4.218)        \\
							\hline
								Year and geographical effects &\multicolumn{2}{c}{Yes}      &\multicolumn{2}{c}{Yes}     &\multicolumn{2}{c}{Yes}        &\multicolumn{2}{c}{Yes}      &\multicolumn{2}{c}{Yes}      &\multicolumn{2}{c}{Yes}     \\
							[1em]
							P-value observable heterogeneity &\multicolumn{2}{c}{ 0.0000 }      &\multicolumn{2}{c}{ 0.0662 }     &\multicolumn{2}{c}{ 0.0142 }        &\multicolumn{2}{c}{ 0.0034  }      &\multicolumn{2}{c}{ 0.3028 }      &\multicolumn{2}{c}{ 0.9556}     \\
							[1em]
							P-value no unobservable heterogeneity &\multicolumn{2}{c}{  0.9980 }      &\multicolumn{2}{c}{ 0.9957 }     &\multicolumn{2}{c}{ 0.8967 }        &\multicolumn{2}{c}{ 0.9988 }      &\multicolumn{2}{c}{1.000 }      &\multicolumn{2}{c}{ 0.9333 }     \\
							[1em]
							Observations &\multicolumn{2}{c}{ 1,357 }      &\multicolumn{2}{c}{ 1,357 }     &\multicolumn{2}{c}{ 1,357 }        &\multicolumn{2}{c}{ 1,357 }      &\multicolumn{2}{c}{ 1,357 }      &\multicolumn{2}{c}{ 1,357 }     \\
							\hline\hline
						\end{tabular}
						\begin{tablenotes}\footnotesize
							\item[] \footnotesize \sym{*} \(p<0.05\), \sym{**} \(p<0.01\), \sym{***} \(p<0.001\).
							\item[] \footnotesize Baseline categories: no education diploma for parents, different occupations (manager, research, technical and academic professors and staff), different industries (finance, public administration, real state, teaching, social health, domestic and others), no insurance, no younger children in the household and for the geographical variable it is living outside the Central Valley in a rural zone.
							\item[] Omitted variables from the table because of their non significant effect: father's years of schooling and indicator variable for father with a mechanic occupation.
							\item[] The reported test statistic for observable heterogeneity tests for the joint significance of all elements in the column "Difference if treated", while the test for unobservable heterogeneity tests if MTEs differ with unobserved costs of treatment. I estimated the model in STATA with the user written command ``mtefe" \citep{andresen2018exploring}.
						\end{tablenotes}
					\end{threeparttable}
				\end{adjustbox}
			\end{table}
		\end{landscape}

		\section{Extensive version of models and their restrictions}
		\subsection{Unitary model}\label{app:restrict_unitary}
		In this section, I present the restrictions from the unitary model \ref{model/unitary}. The model derives two restrictions immediately. First, the ``income pooling" property:
		\begin{equation}\tag{UR1}\label{app:rest_uni1}
			\frac{\mathit{d}m^{p}}{\mathit{d}w^{t}} = \frac{\mathit{d}m^{p}}{\mathit{d}y}, \; \; \; \textrm{when teen works}
		\end{equation}
		which means that a change in $w^{t}$ can only have an income effect on the parents' market supply. On the other hand, if the teen goes to school, this effect is zero:
		\begin{equation}\tag{UR2}\label{app:rest_uni2}
			\frac{\mathit{d}m^{p}}{\mathit{d}w^{t}} = 0, \; \; \; \textrm{when teen goes to school}
		\end{equation}
		Regarding the teen's schooling decision, it depends on the difference between the household's (indirect) utility when he goes to school (denote $V^{s}$) or when she works ($V^{w}$):
		\begin{equation*}
			s^{t} = 1 \iff V^{s}(w^{p},y) \geq V^{w}(w^{p},w^{t}+y)
		\end{equation*} 
		The frontier is characterized by 
		\begin{equation*}
			V^{w}(w^{p},y+\gamma(w^{t}+y)) =  V^{s}(w^{p},y)
		\end{equation*}
		which implies the following condition
		\begin{equation} \tag{UR3}\label{app:rest_uni3}
			\frac{\mathit{d}\gamma}{\mathit{d}w^{p}} = m^{p}_{school} - m^{p}_{work} + m^{p}_{school} \frac{\mathit{d}\gamma}{\mathit{d}y}
		\end{equation}
		The last term on the right-hand side corresponds to a standard income effect. The difference $m^{p}_{school} - m^{p}_{work}$ corresponds to the effect on the teen's schooling decision cost due to a reduction of parents' labor supply. Restrictions \ref{app:rest_uni1}, \ref{app:rest_uni2} and \ref{app:rest_uni3} translate to the empirical model to equations \ref{rest_u1}	and \ref{rest_u2} presented in the paper:
\begin{multicols}{2}
	\noindent
	\begin{equation}\tag{U1}
		\begin{split}
			A_{t} &= A_{y} \\
			a_{t} &= 0
		\end{split}
	\end{equation}
	\columnbreak
	\begin{equation}\tag{U2}
		\begin{split}
			(1+\gamma_{y})&(a_{y}-A_{y}) = 0 \\
			A_{y}\gamma_{p} &= (1+\gamma_{y})a^{*}_{p} - A^{*}_{p}
		\end{split}
	\end{equation}
\end{multicols}

		\subsection{Collective model}\label{app:restrict_collective}
		The collective model \ref{model/hh} provides theoretical restrictions that are imposed in the estimation process. As presented by \citet{blundell2007collective}, the teen's extensive margin decision to go to school or work produces two different outcomes for the household, specifically her parents.
		\\
		If the teen works, her utility is $U^{t}(0,C,f^{K}(h^{p},h^{t}))$, which defines a fix level of utility:
		\begin{equation*}
			U^{t}(0,C,\bar{f}^{K}) = \bar{u}^{t}(w^{p},w^{t},y)
		\end{equation*}
		Solving for consumption $C^{t}$, the optimal consumption is:
		\begin{equation*}
			C^{t} = V^{t}\left[\bar{u}^{t}(w^{p},w^{t},y)\right] = \rho^{t}(w^{p},w^{t},y| \bar{f}^{K})
		\end{equation*}
		where $V^{t}$ is the inverse mapping of $U^{t}(0,\cdot)$. Pareto efficiency is equivalent to the parents' labor supply decision being the solution to the following program:
		\begin{equation}\label{model/parent/work}
			\begin{split}
				\underset{m^{p}, C^{p}}{\max} \; \; &U^{p}(1-m^{p},C^{p},\bar{f}^{K}) \\
				C^{p} &=  w^{p}m^{p} + w^{t} + y - \rho^{t}(w^{p},w^{t},y| \bar{f}^{K}) \\
				\; 0 < &m^{p}\leq 1 
			\end{split}
		\end{equation}
		This generates a labor supply of the form:
		\begin{equation}\label{eq:market/part}
			m^{p}(w^{p},w^{t},y) = M^{p}[w^{p},w^{t} + y - \rho^{t}(w^{p},w^{t},y| \bar{f}^{K})]
		\end{equation}
		where $M^{p}$ is the Marshallian labor supply.

		In the case the teen does not participate in the labor market and instead goes to school, her utility is $U^{t}(1,C,\bar{f}^{K})$ and:
		\begin{equation*}
			U^{t}(1,C,\bar{f}^{K}) = \bar{u}^{t}(w^{p},w^{t},y) = (V^{t})^{-1}[\rho^{t}(w^{p},w^{t},y| \bar{f}^{K})]
		\end{equation*}
		The teen's consumption can be obtained by inverting the previous equation
		\begin{equation*}
			C^{t} = W^{t}[(V^{t})^{-1}(\rho^{t}(w^{p},w^{t},y| \bar{f}^{K}))] = F(\rho^{t}(w^{p},w^{t},y| \bar{f}^{K}))
		\end{equation*}
		where $W^{t}$ is the inverse of the mapping $U^{t}(1,\cdot)$ and $F = W^{t}\circ (V^{t})^{-1} $ is increasing.\\
		The parents' program follows as before and leads to the following labor supply:
		\begin{equation}\label{eq:market/nopart}
			m^{p}(w^{p},w^{t},y) = M^{p}[w^{p}, y^* - F(\rho^{t}(w^{p},w^{t},y| \bar{u}^{K}))]
		\end{equation}
		where $M^{p}$ is the Marshallian labor supply. Lastly, to model the teen's schooling decision, the participation frontier, $\mathit{L}$, is a set of wages and non-labor income bundles $(w^{p},w^{t},y) \in \mathit{L}$, for which the teen is indifferent to attend high school or not. This implies that there exists a reservation wage for the teen $w^{t}_{R}$ such that:
		\begin{equation}\label{eq:reservation}
			\forall (w^{p},w^{t},y) \in \mathit{L}, \; \; \; \rho^{t}(w^{p},w^{t},y| \bar{u}^{K}) - F(\rho^{t}(w^{p},w^{t},y| \bar{u}^{K})) = w^{t}_{R}
		\end{equation}
		Using a shadow wage condition to parametrize $\textit{L}$, as define in equation (\ref{eq:reservation}). The teen works if and only if:
		\begin{equation*}
			w^{t}_{R} > \gamma(w^{p},y)
		\end{equation*}
		for some function $\gamma$ that describes the frontier. 

		The extended version of the empirical specification of the collective model presented in the paper and the derivation of its restrictions goes as follows. First, I focus on the first-stage allocation of the household's non-labor income $y$ through the conditional sharing functions $\rho^{i}(w^{p},w^{t},y| \bar{f}^{K})$. The first-stage household maximization boils down to:
		\begin{subequations}
			\begin{equation}
				\underset{\rho^{p},\rho^{t}, f^{K}}{\max} \; \; \lambda^{p}(w^{p},w^{t},y) v^{p}(w^p,\rho^{p},{f}^{K})  + 
				\lambda^{t}(w^{p},w^{t},y)  v^{t}(w^t,\rho^{t},{f}^{K})
			\end{equation}
			\begin{numcases}{s.t.}
				\lambda^{p}(w^{p},w^{t},y) + 	\lambda^{t}(w^{p},w^{t},y) = 1 \\
				\rho^{p} + \rho^{t} + g^{K}(w^{p},w^{t})f^{K} = y^{*}
			\end{numcases}
		\end{subequations}
		
		Assuming an interior solution with $\mu$ as the Lagrange multiplier, the first order conditions of the associated Lagrangian are:
		\begin{align*}
			&\frac{\partial \mathcal{L}}{\partial \rho^{p}} = \lambda \frac{\textrm{exp}(\theta^{p}_{\rho}w^{p})}{\theta^{p}_{\rho}} \theta^{p}_{\rho} - \mu = 0 \\[10pt]
			&\frac{\partial \mathcal{L}}{\partial \rho^{t}} = (1-\lambda)\theta^{t}_{\rho}  - \mu = 0 \\[10pt] 
			&\frac{\partial \mathcal{L}}{\partial f^{K}} = \lambda \frac{\textrm{exp}(\theta^{p}_{\rho}w^{p})}{\theta^{p}_{\rho}} \frac{\theta^{p}_{K}}{f^K} + (1-\lambda) \frac{\theta^{t}_{K}}{f^K} - \mu g^{K}(w^{p},w^{t})  = 0 \\[10pt] 
			&\frac{\partial \mathcal{L}}{\partial \mu} = y^{*} - \rho^{p} - \rho^{t} - g^{K}(w^{p},w^{t})f^{K}  = 0
		\end{align*}
		From these equations, one obtains the Bowen-Lindahl-Samuelson condition for the optimal provision of public goods inside the household. Then using Roy's identity it is possible to recover the parents' Marshallian labor supply as presented in the paper. The restrictions for the collective model can be separated into two sets. The first set of restrictions comes from the labor participation of the teen. \citet{blundell2007collective} show that equation (\ref{eq:reservation}) has a unique solution for $w^{t}_{R}$ under the following sufficient condition
		\begin{equation*}
			\forall (w^{p},w^{t},y), | [1 - F'(\rho(w^{p},\gamma(w^{p},y),y))]	g^{K}(w^{p},w^{t}) = \frac{1}{\overline{f}^{K} } \bigg(\frac{\theta^{p}_{K}}{\theta^{p}_{\rho}} + \frac{\theta^{t}_{K}}{\theta^{t}_{\rho}}\bigg)(w^{p},w^{t},y) | < 1
		\end{equation*}
		which in my setting implies the following restriction:
		\begin{equation}\label{eq:assumpR}
			|(1 - \theta^{t}_{\rho})\psi_{t}|   < 1
		\end{equation}
		In this case, whenever $m^{p}>0$, $\gamma$ is characterized by the following equation:
		\begin{equation}\label{eq:schooling_front}
			\forall (w^{p},y) \in L, \; \;  \rho(w^{p},\hat{m}^{t}(w^{p},y),y) - F(\rho(w^{p},\gamma(w^{p},y),y))  = \gamma(w^{p},y)
		\end{equation}
		which implies the following two conditions:
		\begin{align}
			\psi_{y} + \gamma_{y}\psi_{t} &= \frac{\gamma_{y}}{1-\theta^{t}_{\rho}} \label{eq:R1}\tag{CR1} \\
			\psi_{p} &= \frac{\gamma_{w^{p}}}{\gamma_{y}}\psi_{y} \label{eq:R2}\tag{CR2}
		\end{align}	
		The next set of restrictions is in the parents' labor supply. If the teen works, for any $(w^{p},w^{t},y) \in P$ such that $m^{p}(w^{p},w^{t},y)>0$:
		\begin{equation*}
			\frac{1 - \psi_{t}}{1-\psi_{y}} = \frac{m^{p}_{w^{t}}}{m^{p}_{y}} = A(w^{p},w^{t},y) 
		\end{equation*}
		In the case the teen goes to school:
		\begin{equation*}
			\frac{- F'\psi_{t}}{1-F'\psi_{y}} = \frac{m^{p}_{w^{t}}}{m^{p}_{y}} = B(w^{p},w^{t},y) 
		\end{equation*}
		Both equations can be rearranged as:
		\begin{align}
			- \psi_{t} + A\psi_{y} = A - 1 \label{eq:R3}\tag{CR3} \\
			- \psi_{t} + B\psi_{y} = \frac{B}{F'} \label{eq:R4}\tag{CR4}
		\end{align}
		Aggregating restrictions (\ref{eq:R1}), (\ref{eq:R2}), (\ref{eq:R3}) and (\ref{eq:R4})  gives the two restrictions \ref{rest_c1} presented in the paper:
\begin{equation}\tag{C1}
	\frac{A_{t} - a_{t}}{A_{y}- a_{y}} =             -\frac{1}{\gamma_{y}}, \; \; \; \frac{A^{*}_{p} - a^{*}_{p}}{A_{y}- a_{y}} =     \frac{\gamma_{p}}{\gamma_{y}}
\end{equation} 	
		
		\subsection{Recovering structural parameters}\label{app:recovering}
		If the data do not reject the teen as a decision maker, I can recover the sharing function of the household as done by \cite{blundell2007collective}. On any point of the frontier, the four restrictions above, (\ref{eq:R1}), (\ref{eq:R2}), (\ref{eq:R3}) and (\ref{eq:R4}), create a non-linear system of  equations in the unknowns $(\psi_{p}, \psi_{t},\psi_{y}, F')$. With some algebra, one obtains the following equation in $F'$:
		\begin{equation*}
			(\gamma_{y}ba - 1 + a - \gamma_{y}b)(F')^{2} + (-b + 1 - 2\gamma_{y}ba + \gamma_{y}a - a)F' + b + \gamma_{y}ba = 0
		\end{equation*} 
		where $a = a(w^{p},y) = A[w^{p},\gamma(w^{p},y),y]$ and likewise for $b$. \citet{blundell2007collective} show that if there is a solution to this quadratic equation that satisfies equation (\ref{eq:assumpR}), then the sharing rule is identified. This solution is such that:
		\begin{equation*}
			F'(\rho^{t}(w^{p},w^{t},y| \bar{u}^{K})) =  \theta_{\rho}^{t}(w^{p},y)
		\end{equation*}
		and $(\psi_{p}, \psi_{t},\psi_{y})$ are recovered with the following equations (rewritten from the restrictions above):
		\begin{align*}
			\psi_{t}[w^{p},\gamma(w^{p},y),y] &= K(w^p,y) = \frac{b}{(a-b)}\bigg(a - 1 -  \frac{a}{\theta_{\rho}^{t}(w^{p},y)}\bigg) \\
			\psi_{p}[w^{p},\gamma(w^{p},y),y] &= L(w^p,y) = \frac{\gamma_{p}}{(a-b)\gamma_{y}}\bigg(a - 1 -  \frac{b}{\theta_{\rho}^{t}(w^{p},y)}\bigg) \\
			\psi_{y}[w^{p},\gamma(w^{p},y),y] &= M(w^p,y) = \frac{1}{(a-b)}\bigg(a - 1 -  \frac{b}{\theta_{\rho}^{t}(w^{p},y)}\bigg)
		\end{align*}
		Then, from the mapping between the structural Marshallian labor supply and its reduced form equation, I recover the last parameters:
		\begin{align*}
			\theta_{\rho}^{p} &=  \frac{A_y}{1-\psi_{y}} \\
			\theta_{w}^{p} &= A_{p}+\theta_{\rho}^{p}\psi_{p}
		\end{align*}
		Lastly, it is important to remind that functions $\rho$ and $F$ are identified up to a constant on the teen schooling participation.

		\section{Wages' imputation}\label{app:imputation}
\begin{figure}[h!]
		\centering
		\caption{Imputation wages}%
		\label{fig:imputation_wages}%
		\begin{subfigure}{.35\linewidth}
			\centering\includegraphics[width=1.3\linewidth]{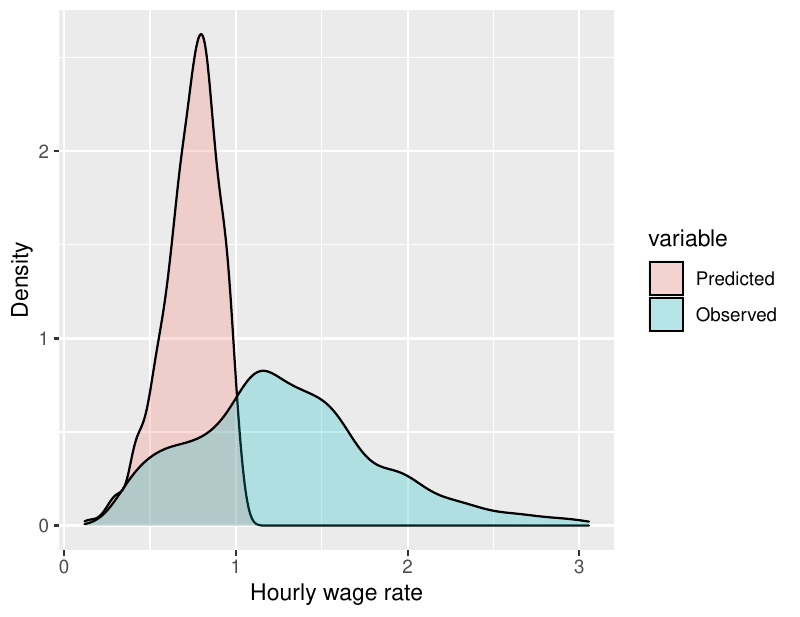}
			\caption{Women.}
		\end{subfigure}\hspace{10mm}
		\begin{subfigure}{.35\linewidth}
			\centering\includegraphics[width=1.3\linewidth]{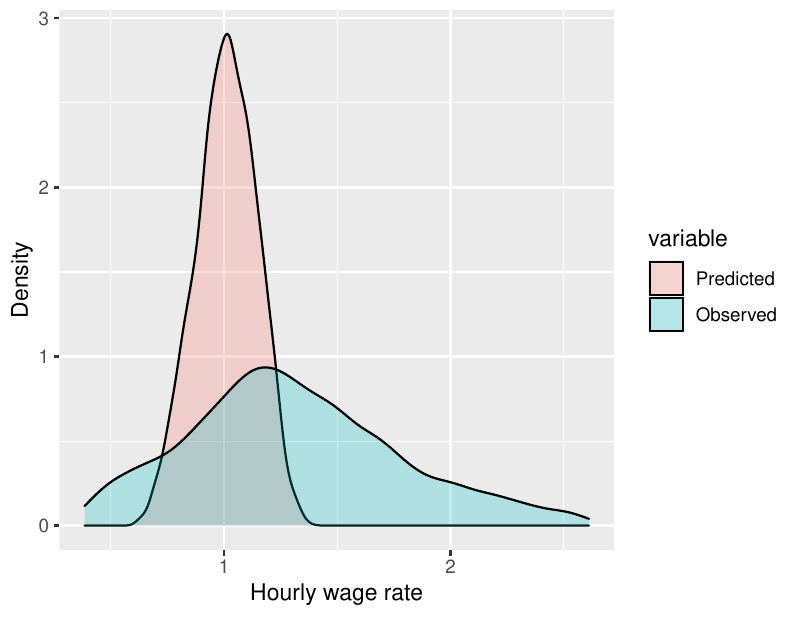}
			\caption{Men.}
		\end{subfigure}
	\end{figure}

		I impute teens' wages using a different sample of the Costa Rican National Household Survey from 2011 to 2019. The sample consists of individuals aged 15 to 25 years old with a school diploma and without a high school diploma who is the child of the head of the household. I impute men's and women's wages separately. The men's sample consists of 15,751 individuals where 4,824 are employed. For the women's sample, there are 1,397 employed women out of 11,810 observations. I impute using a Heckman two-step selection procedure with the following Mince equation for wages:
		\begin{equation}\label{eq:wages}
			w^{i} = \alpha_{0} + \alpha_{1}\textrm{age}^{i} + \alpha_{2}s^{i}  + \mathbf{X}'\mathbf{A} +  u^{i}_{w} 
		\end{equation}
		where $i$ is an individual, $s^{i}$ is years of schooling and $\mathbf{X}$ is a vector of geographical and year effects. For the participation equation, I include as extra covariates demographic variables of the individual and household characteristics such as the number of children in the household, the head of the household's age and years of schooling. Table \ref{tab:imp_men} shows the results of the estimation. For the women, Table \ref{tab:imp_women}  shows the results. Figure \ref{fig:imputation_wages} shows the comparison between the observed and predicted values for both imputations.

		\begin{table}[H] \centering 
			\caption{Men's wage imputation results} 
			\label{tab:imp_men} 
			\begin{adjustbox}{width=11cm,totalheight=23cm}
				\begin{threeparttable}
					\begin{tabular}{@{\extracolsep{5pt}}lD{.}{.}{-3} D{.}{.}{-3} } 
						\\[-1.8ex]\hline 
						\hline \\[-1.8ex] 
						& \multicolumn{2}{c}{\textit{Dependent variable:}} \\ 
						\cline{2-3} 
						\\[-1.8ex] & \multicolumn{1}{c}{Employed} & \multicolumn{1}{c}{Log Hourly wage rate} \\ 
						\hline \\[-1.8ex] 
						Age & 0.268^{***} & 0.058^{***} \\ 
						& (0.005) & (0.011) \\ 
						Years of schooling & &  0.035^{***}  \\ 
						&  &  (0.011)\\ 
						One child in household & -0.033  & \\ 
						&  (0.037) & \\ 
						Two children & -0.033^{*} &  \\ 
						& (0.037) &  \\ 
						Three children & 0.068 &  \\ 
						& (0.046) &  \\ 
						Four or more children & -0.033 &  \\ 
						& (0.053) &  \\ 
						Years of schooling head of household & -0.039^{***} &  \\ 
						& (0.004) &  \\ 
						Age head of household & -0.007^{***} &  \\ 
						& (0.001) &  \\ 
						Constant & -5.029^{***} & -0.425 \\ 
						& (0.112) & (0.292) \\ 
						\hline \\[-1.8ex] 
						Year and geographical effects & \multicolumn{1}{c}{Yes} & \multicolumn{1}{c}{Yes} \\ 
						Observations & \multicolumn{1}{c}{15,751} & \multicolumn{1}{c}{4,824} \\ 
						R$^{2}$ &  & \multicolumn{1}{c}{0.076} \\ 
						Adjusted R$^{2}$ &  & \multicolumn{1}{c}{0.073} \\ 
						Log Likelihood & \multicolumn{1}{c}{-6,833.110} &  \\ 
						Akaike Inf. Crit. & \multicolumn{1}{c}{13,704.220} &  \\ 
						$\rho$ &  & \multicolumn{1}{c}{0.377} \\ 
						Inverse Mills Ratio &  & \multicolumn{1}{c}{0.176$^{***}$  (0.067)} \\ 
						\hline 
						\hline
					\end{tabular}
					\begin{tablenotes}
						\item[] \footnotesize $^{*}$p$<$0.1, $^{**}$p$<$0.05,$^{***}$p$<$0.01.
						\item[] \footnotesize Baseline categories: different occupations (manager, research, technical and academic professors and staff), different industries (finance, public administration, real state, teaching, social health, domestic and others), spouse or another relationship in the household, working in a firm with less than 10 employees and for the geographical variable it is living outside the Central Valley in a rural zone.
					\end{tablenotes}
				\end{threeparttable}
			\end{adjustbox}
		\end{table} 
		
		\begin{table}[H] \centering 
			\caption{Women's wage imputation results} 
			\label{tab:imp_women} 
			\begin{adjustbox}{width=11cm,totalheight=23cm}
				\begin{threeparttable}
					\begin{tabular}{@{\extracolsep{5pt}}lD{.}{.}{-3} D{.}{.}{-3} } 
						\\[-1.8ex]\hline 
						\hline \\[-1.8ex] 
						& \multicolumn{2}{c}{\textit{Dependent variable:}} \\ 
						\cline{2-3} 
						\\[-1.8ex] & \multicolumn{1}{c}{Employed} & \multicolumn{1}{c}{Log Hourly wage rate} \\ 
						\hline \\[-1.8ex] 
						Age & 0.258^{***} & 0.074^{*} \\ 
						& (0.007) & (0.043) \\ 
						Years of schooling &  & 0.025^{**} \\ 
						&  & (0.010) \\ 
						One child in household & 0.064 &  \\ 
						& (0.056) &  \\ 
						Two children & 0.021 &  \\ 
						& (0.058) &  \\ 
						Three children & 0.121^{*} &  \\ 
						& (0.068) &  \\ 
						Four or more children & 0.207^{**} &  \\ 
						& (0.079) &  \\ 
						Years of schooling head of household & -0.012^{**} &  \\ 
						& (0.006) &  \\ 
						Age head of household & -0.004^{*} &  \\ 
						& (0.002) &  \\ 
						Constant & -5.962^{***} & -1.028 \\ 
						& (0.173) & (1.199) \\ 
						\hline \\[-1.8ex] 
						Year and geographical effects & \multicolumn{1}{c}{Yes} & \multicolumn{1}{c}{Yes} \\ 
						Observations & \multicolumn{1}{c}{11,810} & \multicolumn{1}{c}{1,397} \\ 
						R$^{2}$ &  & \multicolumn{1}{c}{0.050} \\ 
						Adjusted R$^{2}$ &  & \multicolumn{1}{c}{0.039} \\ 
						Log Likelihood & \multicolumn{1}{c}{-2,872.722} &  \\ 
						Akaike Inf. Crit. & \multicolumn{1}{c}{5,783.444} &  \\ 
						$\rho$ &  & \multicolumn{1}{c}{0.532} \\ 
						Inverse Mills Ratio &  & \multicolumn{1}{c}{0.312  (0.224)} \\  
						\hline 
						\hline
					\end{tabular}
					\begin{tablenotes}
						\item[] \footnotesize $^{*}$p$<$0.1, $^{**}$p$<$0.05,$^{***}$p$<$0.01.
						\item[] \footnotesize Baseline categories: different occupations (manager, research, technical and academic professors and staff), different industries (finance, public administration, real state, teaching, social health, domestic and others), spouse or another relationship in the household, working in a firm with less than 10 employees and for the geographical variable it is living outside the Central Valley in a rural zone.
					\end{tablenotes}
				\end{threeparttable}
			\end{adjustbox}
		\end{table}

		\section{Control function approach}\label{app:control_function}
		For non-labor income, I define as the difference between the household's total income and its labor income. The sample I use is the same as presented in the paper. To obtain the residuals I use in the control function approach I estimate the following regression:
		\begin{equation}\label{eq:nonlaborincome}
			y = \; \alpha_{0}^{y} + \alpha_{1}\textrm{IV}_{i} + \alpha_{2}\textrm{IV}_{i}^2  + \mathbf{X}'\mathbf{A} + u_{yi}
		\end{equation}
		where $i$ is a household, $IV$ is an instrument and $X$ is a vector of covariates containing demographics and geographical and year effects. The instrument I use is the amount the household receives if it benefits from \textit{Avancemos}. Table \ref{tab:imputation_non_labor} show the results of the estimation.
		
		\begin{table}[H] \centering 
			\caption{Non-labor income imputation results} 
			\label{tab:imputation_non_labor} 
			\begin{adjustbox}{width=11cm,totalheight=23cm}
				\begin{threeparttable}
					\begin{tabular}{@{\extracolsep{5pt}}lD{.}{.}{-3} } 
						\\[-1.8ex]\hline 
						\hline \\[-1.8ex] 
						& \multicolumn{1}{c}{\textit{Dependent variable:}} \\ 
						\cline{2-2} 
						\\[-1.8ex] & \multicolumn{1}{c}{non-labor income} \\ 
						\hline \\[-1.8ex] 
						IV & -1.085^{*} \\ 
						& (0.591) \\ 
						IV square& 0.094 \\ 
						& (0.061) \\ 
						Age teen & 1.617^{***} \\ 
						& (0.432) \\ 
						Female teen & -2.274^{**} \\ 
						& (1.124) \\ 
						Age father & 0.075 \\ 
						& (0.102) \\ 
						Age mother & -0.281^{**} \\ 
						& (0.127) \\ 
						Father's years of schooling & 0.307 \\ 
						& (0.285) \\ 
						Mother's years of schooling & 0.985^{***} \\ 
						& (0.293) \\ 
						Father high school diploma or more & 4.902^{**} \\ 
						& (2.308) \\ 
						Mother high school diploma or more & 7.269^{***} \\ 
						& (2.365) \\ 
						One kid & -0.207 \\ 
						& (1.469) \\ 
						Two or more kids & 3.520^{**} \\ 
						& (1.685) \\ 
						Constant & 24.132^{***} \\ 
						& (8.599) \\ 
						\hline \\[-1.8ex] 
						Year and geographical effects & \multicolumn{1}{c}{Yes} \\ 
						Observations & \multicolumn{1}{c}{2,447} \\ 
						Log Likelihood & \multicolumn{1}{c}{-11,550.520} \\ 
						Akaike Inf. Crit. & \multicolumn{1}{c}{23,149.040} \\ 
						\hline 
						\hline
					\end{tabular}
					\begin{tablenotes}
						\item[] \footnotesize $^{*}$p$<$0.1, $^{**}$p$<$0.05,$^{***}$p$<$0.01.
						\item[] \footnotesize Baseline categories: positive rent income, household of 3 members (father, mother and teen) and for the geographical variable it is living outside the Central Valley in a rural zone.
					\end{tablenotes}
				\end{threeparttable}
			\end{adjustbox}
		\end{table}

	\section{Structural results}\label{app:estimates}
		
	In this section I show the results of the structural model. As explained in the paper, I estimate an unrestricted model, the unitary model and the collective model. I show three the results for each estimation for the sons and daughters samples.

\begin{table}[H]
	\centering
	\caption{Estimates unrestricted model - daughters}\label{tab:unr_daughters}
		\begin{threeparttable}
	\begin{tabular}{lcccccccc}\hline 	\hline 
		& \multicolumn{5}{c}{\textbf{Parents weekly labor hours}} &       & \multicolumn{2}{c}{\textbf{Teen school decision}} \\
		\hline 
		& \multicolumn{2}{c}{Teen no school} &       & \multicolumn{2}{c}{Teen school} &       &       &  \\
	\hline 
		& \textbf{Coef} & \textbf{SE} &       & \textbf{Coef} & \textbf{SE} &       & \textbf{Coef} & \textbf{SE} \\ \hline 
		Hourly wage teen & 9.478 & (8.179) &       & 11.979 & (6.610) &       & -1.336 & (0.262) \\
		Hourly wage parents & -16.763 & (5.995) &       & -15.406 & (2.512) &       & 0.479 & (0.131) \\
		Non-labor income & 26.685 & (19.408) &       & 38.098 & (7.175) &       & 0.892 & (0.390) \\
		Intercept & 34.042 & (20.134) &       & 22.761 & (7.482) &       & 0.321 & (0.419) \\
		Control function  & -18.759 & (18.842) &       & -22.988 & (7.260) &       & -0.860 & (0.390) \\
		\midrule
		\cmidrule{1-6}\cmidrule{8-9}    \textbf{N} & \multicolumn{2}{c}{148} &       & \multicolumn{2}{c}{942} &       & \multicolumn{2}{c}{1,090} \\
		\hline 	\hline 
	\end{tabular}%
\begin{tablenotes}
	\item[] \footnotesize The variable "Control function" refers to the residuals of the regression on non-labor income.
	\item[] \footnotesize Missing values for some variables are due to the restrictions impose in the estimation from the model.
\end{tablenotes}
\end{threeparttable}
\end{table}%

\begin{table}[H]
	\centering
	\caption{Estimates unitary model - daughters}\label{tab:unitary_daughters}
	\begin{threeparttable}
	\begin{tabular}{lcccccccc}	\hline 	\hline 
		& \multicolumn{5}{c}{\textbf{Parents weekly labor hours}} &       & \multicolumn{2}{c}{\textbf{Teen school decision}} \\
		\hline 
		& \multicolumn{2}{c}{Teen no school} &       & \multicolumn{2}{c}{Teen school} &       &       &  \\
		\hline 
		& \textbf{Coef} & \textbf{SE} &       & \textbf{Coef} & \textbf{SE} &       & \textbf{Coef} & \textbf{SE} \\ \hline 
		Hourly wage teen & 29.097 & (5.517) &       &       &       &       & -1.451 & (0.245) \\
		Hourly wage parents & -26.282 & (4.400) &       &       &       &       & 0.439 & (0.130) \\
		Non-labor income &       &       &       &       &       &       & 0.767 & (0.387) \\
		Intercept & 23.238 & (9.046) &       & 36.268 & (5.795) &       & 0.539 & (0.397) \\
		Control function & -20.330 & (7.573) &       & -15.155 & (5.746) &       & -0.724 & (0.386) \\
		\cmidrule{1-6}\cmidrule{8-9}    \textbf{N} & \multicolumn{2}{c}{148} &       & \multicolumn{2}{c}{942} &       & \multicolumn{2}{c}{1,090} \\
			\hline 	\hline 
		\end{tabular}%
		\begin{tablenotes}
			\item[] \footnotesize The variable "Control function" refers to the residuals of the regression on non-labor income.
			\item[] \footnotesize Missing values for some variables are due to the restrictions impose in the estimation from the model.
		\end{tablenotes}
	\end{threeparttable}
\end{table}%

\begin{table}[H]
	\centering
	\caption{Estimates collective model - daughters}\label{tab:collective_daughters}
	\begin{threeparttable}
	\begin{tabular}{lcccccccc}	\hline 	\hline 
		& \multicolumn{5}{c}{\textbf{Parents weekly labor hours}} &       & \multicolumn{2}{c}{\textbf{Teen school decision}} \\
		\hline 
		& \multicolumn{2}{c}{Teen no school} &       & \multicolumn{2}{c}{Teen school} &       &       &  \\
		\hline 
		& \textbf{Coef} & \textbf{SE} &       & \textbf{Coef} & \textbf{SE} &       & \textbf{Coef} & \textbf{SE} \\ \hline 
		Hourly wage teen &       &       &       & 7.211 & (7.783) &       & -1.183 & (0.420) \\
		Hourly wage parents &       &       &       & -14.606 & (2.566) &       & 0.446 & (0.156) \\
		Non-labor income & 24.450 & (23.936) &       & 36.279 & (7.350) &       & 1.269 & (0.647) \\
		Intercept & 32.456 & (21.782) &       & 26.682 & (7.069) &       & -0.155 & (0.813) \\
		Control function & -14.984 & (23.217) &       & -21.461 & (7.562) &       & -1.234 & (0.637) \\
		\cmidrule{1-6}\cmidrule{8-9}    \textbf{N} & \multicolumn{2}{c}{148} &       & \multicolumn{2}{c}{942} &       & \multicolumn{2}{c}{1,090} \\
			\hline 	\hline 
		\end{tabular}%
		\begin{tablenotes}
			\item[] \footnotesize The variable "Control function" refers to the residuals of the regression on non-labor income.
			\item[] \footnotesize Missing values for some variables are due to the restrictions impose in the estimation from the model.
		\end{tablenotes}
	\end{threeparttable}
\end{table}%

\begin{table}[H]
	\centering
	\caption{Estimates unrestricted model - sons}\label{tab:unr_sons}
	\begin{threeparttable}
		\begin{tabular}{lcccccccc}\hline 	\hline 
		& \multicolumn{5}{c}{\textbf{Parents weekly labor hours}} &       & \multicolumn{2}{c}{\textbf{Teen school decision}} \\
		\hline 
		& \multicolumn{2}{c}{Teen no school} &       & \multicolumn{2}{c}{Teen school} &       &       &  \\
		\hline 
		& \textbf{Coef} & \textbf{SE} &       & \textbf{Coef} & \textbf{SE} &       & \textbf{Coef} & \textbf{SE} \\ \hline 
		Hourly wage teen & 3.461 & (4.468) &       & 24.535 & (9.524) &       & -1.804 & (0.220) \\
		Hourly wage parents & -5.946 & (3.841) &       & -12.117 & (2.388) &       & 0.745 & (0.109) \\
		Non-labor income & 21.169 & (12.673) &       & 21.504 & (6.160) &       & 1.019 & (0.290) \\
		Intercept & 33.432 & (14.387) &       & 23.054 & (9.595) &       & 0.503 & (0.354) \\
		Control function & -12.737 & (12.605) &       & -12.644 & (6.335) &       & -1.195 & (0.295) \\
		\cmidrule{1-6}\cmidrule{8-9}    \textbf{N} & \multicolumn{2}{c}{342} &       & \multicolumn{2}{c}{1,015} &       & \multicolumn{2}{c}{1,357} \\
			\hline 	\hline 
		\end{tabular}%
		\begin{tablenotes}
			\item[] \footnotesize The variable "Control function" refers to the residuals of the regression on non-labor income.
			\item[] \footnotesize Missing values for some variables are due to the restrictions impose in the estimation from the model.
		\end{tablenotes}
	\end{threeparttable}
\end{table}%

\begin{table}[H]
	\centering
	\caption{Estimates unitary model - sons}\label{tab:unitary_sons}
	\begin{threeparttable}
		\begin{tabular}{lcccccccc}	\hline 	\hline 
		& \multicolumn{5}{c}{\textbf{Parents weekly labor hours}} &       & \multicolumn{2}{c}{\textbf{Teen school decision}} \\
		\hline 
		& \multicolumn{2}{c}{Teen no school} &       & \multicolumn{2}{c}{Teen school} &       &       &  \\
		\hline 
		& \textbf{Coef} & \textbf{SE} &       & \textbf{Coef} & \textbf{SE} &       & \textbf{Coef} & \textbf{SE} \\ \hline 
			Hourly wage teen & 17.480 & (3.556) &       &       &       &       & -1.910 & (0.214) \\
			Hourly wage parents & -15.322 & (3.037) &       &       &       &       & 0.707 & (0.109) \\
			Non-labor income &       &       &       &       &       &       & 0.933 & (0.294) \\
			Intercept & 26.650 & (7.376) &       & 45.869 & (4.149) &       & 0.716 & (0.351) \\
			Control function & -9.498 & (4.524) &       & -10.096 & (3.963) &       & -1.102 & (0.300) \\
			\cmidrule{1-6}\cmidrule{8-9}    \textbf{N} & \multicolumn{2}{c}{342} &       & \multicolumn{2}{c}{1,015} &       & \multicolumn{2}{c}{1,357} \\
			\hline 	\hline 
		\end{tabular}%
		\begin{tablenotes}
			\item[] \footnotesize The variable "Control function" refers to the residuals of the regression on non-labor income.
			\item[] \footnotesize Missing values for some variables are due to the restrictions impose in the estimation from the model.
		\end{tablenotes}
	\end{threeparttable}
\end{table}%

\begin{table}[H]
	\centering
	\caption{Estimates collective model - sons}\label{tab:collective_sons}
	\begin{threeparttable}
		\begin{tabular}{lcccccccc}	\hline 	\hline 
			& \multicolumn{5}{c}{\textbf{Parents weekly labor hours}} &       & \multicolumn{2}{c}{\textbf{Teen school decision}} \\
			\hline 
			& \multicolumn{2}{c}{Teen no school} &       & \multicolumn{2}{c}{Teen school} &       &       &  \\
			\hline 
			& \textbf{Coef} & \textbf{SE} &       & \textbf{Coef} & \textbf{SE} &       & \textbf{Coef} & \textbf{SE} \\ \hline 
			Hourly wage teen &       &       &       & 21.117 & (7.528) &       & -1.846 & (0.218) \\
			Hourly wage parents &       &       &       & -12.018 & (2.238) &       & 0.747 & (0.108) \\
			Non-labor income & 26.709 & (6.478) &       & 17.911 & (5.555) &       & 0.911 & (0.288) \\
			Intercept & 27.014 & (7.752) &       & 29.970 & (7.902) &       & 0.665 & (0.351) \\
			Control function & -18.250 & (6.873) &       & -9.011 & (5.797) &       & -1.086 & (0.294) \\
			\cmidrule{1-6}\cmidrule{8-9}    \textbf{N} & \multicolumn{2}{c}{342} &       & \multicolumn{2}{c}{1,015} &       & \multicolumn{2}{c}{1,357} \\
			\hline 	\hline 
		\end{tabular}%
		\begin{tablenotes}
			\item[] \footnotesize The variable "Control function" refers to the residuals of the regression on non-labor income.
			\item[] \footnotesize Missing values for some variables are due to the restrictions impose in the estimation from the model.
		\end{tablenotes}
	\end{threeparttable}
\end{table}%

\begin{table}[H]
	\centering
	\caption{Estimates home production}\label{tab:home_production}
	\begin{threeparttable}
		\begin{tabular}{lccccc}	\hline 	\hline 
			\multicolumn{6}{c}{\textbf{Household production function}} \\
			\midrule
			& \textbf{Coef} & \textbf{SE} &       & \textbf{Coef} & \textbf{SE} \\
			\midrule
			\textit{\textbf{Production parameter}} &       &       &       &       &  \\
			Teen  & 0.045 & (0.002) &       & 0.038 & (0.002) \\
			Parents & 0.955 & (0.002) &       & 0.962 & (0.002) \\
			&       &       &       &       &  \\
			Sample & \multicolumn{2}{c}{Daughters} &       & \multicolumn{2}{c}{Sons} \\
			\midrule
			\textbf{N} & \multicolumn{2}{c}{1,090} &       & \multicolumn{2}{c}{1,357} \\
			\hline 	\hline 
		\end{tabular}%
		\begin{tablenotes}
			\item[] \footnotesize Standard errors computed with the Delta Method.
			\item[] \footnotesize For the daughters' sample, the estimates are obtain from an OLS regression. For the sons, the estimates are from the maximization of the likelihood presented in the paper. 
		\end{tablenotes}
	\end{threeparttable}
\end{table}%

\end{document}